\documentclass[journal]{IEEEtran}

\usepackage{cite} 

\usepackage[caption=false,font=footnotesize]{subfig}	
\usepackage{graphicx}

\usepackage{algorithm}
\usepackage{algorithmicx}
\usepackage{algpseudocode}

\usepackage{amsmath,amsfonts,amssymb,amsthm}
\usepackage{mathtools}
\usepackage{mathrsfs}	
\usepackage{subdepth}	

\usepackage{array}	
\usepackage{tabularx}   

\usepackage{hyperref}
\usepackage[sort,compress,capitalise]{cleveref}	
\usepackage{url}

\usepackage[dvipsnames]{xcolor} 
\usepackage{xspace} 

\usepackage{tikz}
\usetikzlibrary{babel}  
\usepackage[straightvoltages,nooldvoltagedirection]{circuitikz}

{
\theoremstyle{plain}
\newtheorem{Hypothesis}{Hypothesis}

\newtheorem{Lemma}{Lemma}

\newtheorem{Definition}{Definition}
}

{
\theoremstyle{definition}

}

\crefname{Hypothesis}{Hyp.}{Hyps.}
\Crefname{Hypothesis}{Hyp.}{Hyps.}

\crefname{Lemma}{Lemma}{Lemmata}
\Crefname{Lemma}{Lemma}{Lemmata}

\crefname{Definition}{Def.}{Defs.}
\Crefname{Definition}{Def.}{Defs.}

\newcommand{\CIGRE}{CIGR{\'E}\xspace}

\newcommand{\DFT}[1][]{DFT\xspace}  

\newcommand{\ADC}[1][]{ADC#1\xspace}  
\newcommand{\DAC}[1][]{DAC#1\xspace}  

\newcommand{\LPF}[1][]{LPF#1\xspace}  

\newcommand{\LTI}{LTI\xspace}	
\newcommand{\LTP}{LTP\xspace}	

\newcommand{\TE}[1][]{\textup{TE#1}\xspace}   
\newcommand{\NE}[1][]{\textup{NE#1}\xspace}   

\newcommand{\DAE}[1][]{DAE#1\xspace}		

\newcommand{\MNA}{MNA\xspace}		
\newcommand{\MANA}{MANA\xspace}		

\newcommand{\AC}{AC\xspace} 
\newcommand{\DC}{DC\xspace}	

\newcommand{\ADN}[1][]{ADN#1\xspace}	
\newcommand{\DER}[1][]{DER#1\xspace}    
\newcommand{\CIDER}[1][]{CI\DER[#1]}    

\newcommand{\HA}{HA\xspace}		
\newcommand{\DHA}{D\HA}			
\newcommand{\IHA}{I\HA}			
\newcommand{\HPF}{HPF\xspace}	

\newcommand{\EMTP}{EMTP\xspace}		
\newcommand{\SPICE}{SPICE\xspace}	

\newcommand{\PLL}[1][]{PLL#1\xspace}		

\newcommand{\RealNum}{\mathbb{R}}   
\newcommand{\CompNum}{\mathbb{C}}   

\newcommand{\Real}[1]{\Re\left\{#1\right\}}

\newcommand{\Exp}[1]{\exp\left(#1\right)}

\newcommand{\Set}[1]{\mathcal{#1}}
\newcommand{\Card}[1]{\left|#1\right|}

\newcommand{\diag}{\operatorname{diag}}

\newcommand{\col}{\operatorname{col}}

\newcommand{\Graph}[1]{\mathfrak{#1}}

\newcommand{\grid}{\gamma}
\newcommand{\pwr}{\pi}
\newcommand{\ctrl}{\kappa}
\newcommand{\act}{\alpha}
\newcommand{\fltr}{\varphi}
\newcommand{\trafo}{\tau}
\newcommand{\refr}{\rho}
\newcommand{\spt}{\sigma}

\newcommand{\V}{\mathbf{V}}	
\newcommand{\I}{\mathbf{I}}	
\newcommand{\Z}{\mathbf{Z}}     
\newcommand{\Y}{\mathbf{Y}}     
\newcommand{\HG}{\mathbf{H}}    

\newcommand{\ground}{\Set{G}}
\newcommand{\phases}{\Set{P}}
\newcommand{\nodes}{\Set{N}}
\newcommand{\branches}{\Set{L}}
\newcommand{\shunts}{\Set{T}}

\newcommand{\branchgraph}{\Graph{B}}

\newcommand{\phsA}{\texttt{A}}
\newcommand{\phsB}{\texttt{B}}
\newcommand{\phsC}{\texttt{C}}

\newcommand{\cmpD}{\texttt{D}}
\newcommand{\cmpQ}{\texttt{Q}}

\newcommand{\harmonics}{\Set{H}}
\newcommand{\formers}{\Set{S}}
\newcommand{\followers}{\Set{R}}

\begin{document}

\title{\huge{%
    Harmonic Power-Flow Study of Polyphase Grids\\
	with Converter-Interfaced Distributed Energy Resources,\\
	Part I: Modelling Framework and Algorithm
}}

\author{%
	Andreas~Martin~Kettner,~\IEEEmembership{Member,~IEEE},
	Lorenzo~Reyes-Chamorro,~\IEEEmembership{Senior~Member,~IEEE},\\
	Johanna~Kristin~Maria~Becker,~\IEEEmembership{Member,~IEEE},
	Zhixiang~Zou,~\IEEEmembership{Member,~IEEE},\\
	Marco~Liserre,~\IEEEmembership{Fellow,~IEEE},
	and~Mario~Paolone,~\IEEEmembership{Senior~Member,~IEEE}%
	\thanks{A. Kettner is with PSI NEPLAN AG, 8700 Küsnacht, Switzerland (E-mail: andreas.kettner@neplan.ch).}%
	\thanks{L. Reyes-Chamorro is with the Facultad de Ciencias de la Ingenier{\'i}a at the Universidad Austral de Chile (UACh) in CL-5111187 Valdivia, Chile (E-mail: lorenzo.reyes@uach.cl).}%
	\thanks{J. Becker, and M. Paolone are with the Distributed Electrical Systems Laboratory at the {\'E}cole Polytechnique F{\'e}d{\'e}rale de Lausanne (EPFL) in CH-1015 Lausanne, Switzerland (E-mail: \{johanna.becker, mario.paolone\}@epfl.ch).}%
	\thanks{Z. Zou is with the School of Electrical Engineering, Southeast University, in PRC-210096 Nanjing, China (E-mail: zzou@seu.edu.cn).}%
	\thanks{M. Liserre is with the Chair of Power Electronics at the Christian-Albrechts-Universit{\"a}t zu Kiel (CAU) in DE-24143 Kiel, Germany (E-mail: ml@tf.uni-kiel.de).}%
	\thanks{This work was funded by the Schweizerischer Nationalfonds (SNF, Swiss National Science Foundation) via the National Research Programme NRP~70 ``Energy Turnaround'' (NRP 70 "Energy Turnaround" (projects nr. 173661 and 197060) and by the Deutsche Forschungsgemeinschaft (DFG, German Research Foundation) via the Priority Programme DFG~SPP~1984 ``Hybrid and Multimodal Energy Systems'' (project nr. 359982322).}%
}

\maketitle







\begin{abstract}
    Power distribution systems are experiencing a large-scale integration of \emph{Converter-Interfaced Distributed Energy Resources} (\CIDER[s]).
    This complicates the analysis and mitigation of harmonics, whose creation and propagation are facilitated by the interactions of converters and their controllers through the grid.
    In this paper, a method for the calculation of the so-called \emph{Harmonic Power-Flow} (\HPF) in three-phase grids with \CIDER[s] is proposed.
    The distinguishing feature of this \HPF method is the generic and modular representation of the system components.
    Notably, as opposed to most of the existing approaches, the coupling between harmonics is explicitly considered.
    The \HPF problem is formulated by combining the hybrid nodal equations of the grid with the closed-loop transfer functions of the \CIDER[s], and solved using the Newton-Raphson method.
    The grid components are characterized by compound electrical parameters, which allow to represent both transposed or non-transposed lines.
    The \CIDER[s] are represented by modular linear time-periodic systems, which allows to treat both grid-forming and grid-following control laws.
    The method's accuracy and computational efficiency are confirmed via time-domain simulations of the \CIGRE low-voltage benchmark microgrid.
    This paper is divided in two parts, which focus on the development (Part~I) and the validation (Part~II) of the proposed method.%
\end{abstract}


\begin{IEEEkeywords}
	Distributed energy resources,
	harmonic power-flow study,
	polyphase power systems,
	power electronic converters,
	unbalanced power grids.
\end{IEEEkeywords}



\section*{Nomenclature}

\begin{center}
    
\begin{tabularx}{\columnwidth}{p{3cm}p{5cm}}
    \hline
    \multicolumn{2}{c}{Grid Model}\\
    \hline
\end{tabularx}
\begin{IEEEdescription}[\IEEEusemathlabelsep\IEEEsetlabelwidth{$\formers\cup\followers$}]
    \item[$g\in\ground$] 
        The ground node ($\ground\coloneqq\{0\}$)
    \item[$p\in\phases$] 
        A phase terminal ($\phases\coloneqq\{\phsA,\phsB,\phsC\}$)
    \item[$n\in\nodes$]
        A three-phase node ($\nodes\coloneqq\{1,...,N\}$)
    \item[$\V_{n}$] 
        The phasors of the nodal voltages at node $n\in\nodes$
    \item[$\I_{n}$]
        The phasors of the injected currents at node $n\in\nodes$
    \item[$\ell\in\branches$] 
        A branch element ($\ell=(m,n):~m,n\in\nodes$)
    \item[$\Z_{\ell}$] 
        A compound branch impedance at $\ell\in\branches$
    \item[$\I_{\ell}$]
        The phasors of the current flows through $\ell\in\branches$
    \item[$t\in\shunts$] 
        A shunt element ($t=(n,g):~n\in\nodes,~g\in\ground$)
    \item[$\Y_{t}$] 
        A compound shunt admittance at $t\in\shunts$
    \item[$\I_{\ell}$]
        The phasors of the current flows through $t\in\shunts$
    \item[$\branchgraph$] 
        The branch graph ($\branchgraph\coloneqq(\nodes,\branches)$)
    \item[$\mathbf{A}_{\branchgraph}$] 
        The three-phase branch incidence matrix
    \item[$\Y$]
        The compound nodal admittance matrix
    \item[$\formers\cup\followers$]
        A partition of $\nodes$ ($\nodes=\formers\cup\followers$, $\formers\cap\followers=\emptyset$)
    \item[$\I_{\formers}$]
        The phasors of the injected currents at all $s\in\formers$
    \item[$\V_{\followers}$]
        The phasors of the nodal voltages at all $r\in\followers$
    \item[$\Y_{\formers\times\followers}$] 
        The block of $\Y$ linking $\I_{\formers}$ and $\V_{\followers}$
    \item[$\HG$] 
        The compound nodal hybrid matrix (w.r.t. $\formers,\followers$)
    \item[$\HG_{\formers\times\followers}$] 
        The block of $\HG$ linking $\V_{\formers}$ and $\V_{\followers}$
    \item[$f$] 
        An arbitrary frequency
    \item[$f_1$]
        The fundamental frequency ($f_{1}\coloneqq\frac{1}{T}$)
    \item[$h\in\harmonics$] 
        A harmonic order ($\harmonics\coloneqq\{-h_{\max},\ldots,h_{\max}\}$)
    \item[$f_{h}$]
        The harmonic frequency of order $h$ ($f_{h}\coloneqq h\cdot f_{1}$)
    \item[$\hat{\V}_{\formers}$] 
        The column vector composed of the Fourier coefficients of $\V_{\formers}$
    \item[$\hat{\HG}_{\formers\times\followers}$] 
        The Toeplitz matrix of the Fourier coefficients of $\HG_{\formers\times\followers}$ (i.e., $\HG_{\formers\times\followers}(f)$ evaluated at $f=f_{h}$)
\end{IEEEdescription}
\hrule

\end{center}

\begin{center}
    
\begin{tabularx}{\columnwidth}{p{3cm}p{5cm}}
    \hline
    \multicolumn{2}{c}{\CIDER Model}\\
    \hline
\end{tabularx}
\begin{IEEEdescription}[\IEEEusemathlabelsep\IEEEsetlabelwidth{$\formers\cup\followers$}]
    \item[$\grid$]
        The power grid
    \item[$\pwr$]
        The power hardware of a \CIDER
    \item[$\ctrl$]
        The control software of a \CIDER
    \item[$\act$]
        The actuator of a \CIDER
    \item[$\refr$] 
        The reference calculation of a \CIDER
    \item[$\spt$] 
        The setpoint of a \CIDER
    \item[$\lambda$]
        A stage inside the cascaded structure of a \CIDER ($\lambda\in\{1,\ldots,\Lambda\}$)
    \item[$\varphi_{\lambda}$] 
        The filter element associated with stage $\lambda$
    \item[$\ctrl_{\lambda}$] 
        The controller element associated with stage $\lambda$
    \item[$\mathbf{x}(t)$]
        The state vector of a state-space model
    \item[$\mathbf{u}(t)$]
        The input vector of a state-space model
    \item[$\mathbf{y}(t)$]
        The output vector of a state-space model
    \item[$\mathbf{w}(t)$]
        The disturbance vector of a state-space model
    \item[$\mathbf{A}(t)$]
        The system matrix of an \LTP system
    \item[$\mathbf{B}(t)$]
        The input matrix of an \LTP system
    \item[$\mathbf{C}(t)$]
        The output matrix of an \LTP system
    \item[$\mathbf{D}(t)$]
        The feed-through matrix of an \LTP system
    \item[$\mathbf{E}(t)$]
        The input disturbance matrix of an \LTP system
    \item[$\mathbf{F}(t)$]
        The output disturbance matrix of an \LTP system
    \item[$\trafo_{\ctrl|\pwr}$]
        A change of reference frame from $\pwr$ to $\ctrl$
    \item[$\mathbf{T}_{\ctrl|\pwr}(t)$] 
        The \LTP matrix which describes $\trafo_{\ctrl|\pwr}$
   \item[$\mathbf{X}_{h}$] 
        The Fourier coefficients of $\mathbf{x}(t)$ ($h\in\harmonics$)
    \item[$\hat{\mathbf{X}}$] 
        The column vector composed of the $\mathbf{X}_{h}$
    \item[$\mathbf{A}_{h}$] 
        The Fourier coefficients of $\mathbf{A}(t)$ ($h\in\harmonics$)
    \item[$\hat{\mathbf{A}}$] 
        The Toeplitz matrix composed of the $\mathbf{A}_{h}$
    \item[$\Hat{\mathbf{G}}$] 
        The harmonic-domain closed-loop gain
    \item[$\partial_{\grid}$] 
        The partial derivative w.r.t. $\Hat{\mathbf{W}}_{\grid}$
\end{IEEEdescription}
\hrule

\end{center}

\begin{center}
    
\begin{tabularx}{\columnwidth}{p{3cm}p{5cm}}
    \hline
    \multicolumn{2}{c}{\HPF Study}\\
    \hline
\end{tabularx}
\begin{IEEEdescription}[\IEEEusemathlabelsep\IEEEsetlabelwidth{$\formers\cup\followers$}]
    \item[$\formers$]
        The nodes with grid-forming \CIDER[s]
    \item[$\followers$] 
        The nodes with grid-following \CIDER[s]
    \item[$\Delta\Hat{\mathbf{V}}_{\formers}$]
        The mismatch equations w.r.t. $\Hat{\mathbf{V}}_{\formers}$
    \item[$\Delta\Hat{\mathbf{I}}_{\followers}$] 
        The mismatch equations w.r.t. $\Hat{\mathbf{I}}_{\followers}$
    \item[$\partial_{\formers}$]
        The partial derivative w.r.t. $\Hat{\mathbf{I}}_{\formers}$
    \item[$\partial_{\followers}$] 
        The partial derivative w.r.t. $\Hat{\mathbf{V}}_{\followers}$
\end{IEEEdescription}
\hrule

\end{center}

\section{Introduction}
\label{sec:intro}

%
%
%



\IEEEPARstart{P}{ower} distribution systems are undergoing a large-scale integration of distributed energy resources, such as renewable generators, energy storage systems, and modern loads.
Typically, these resources are interfaced with the grid via power electronic converters.
The controllability of such \emph{Converter-Interfaced Distributed Energy Resources} (\CIDER[s]) is a crucial asset for power-system operation \cite{Jrn:PSE:PEC:2004:Blaabjerg}.
Moreover, thanks to recent advances in power-system instrumentation and state estimation (e.g., \cite{Bk:PSE:2016:Milano}), real-time situational awareness is nowadays available for power distributions systems.
The deployment of such automation technology is contributing to the development of \emph{Active Distribution Networks} (\ADN[s]), whose power flows can be regulated, mainly by controlling the \CIDER[s] \cite{Rep:PSE:C:2011:CIGRE}.
However, the presence of large numbers of \CIDER[s] can jeopardize the stability of the system.
Therefore, it is vital to first understand the causes of instabilities, and then apply this knowledge to design robust controllers.



Recently, several standardization committees have worked on the classification, modeling, and analysis of stability issues in \ADN[s] (e.g., \cite{Rep:PSE:SA:2018:Canizares,Rep:PSE:SA:2020:Hatziargyriou}).
The instabilities observed in such systems are related to the transfer or balance of power in the grid, or interactions between the resources.
Due to the prevalence of \CIDER[s], \emph{converter} a.k.a \emph{harmonic stability} is of particular importance \cite{Rep:PSE:SA:2018:Canizares}.
Namely, the interaction of \CIDER[s] through the grid can lead to unstable oscillations at harmonic frequencies (e.g., \cite{Jrn:PSE:PEC:2004:Enslin}).

Whether a particular subsystem is a source of excessive harmonics can be detected by a variety of empirical indicators \cite{Jrn:PSE:SSA:2017:Safargholi:Part1,Jrn:PSE:SSA:2017:Safargholi:Part2}.
However, in order to design controllers which are robust w.r.t. harmonic instability, the creation and propagation of harmonics must be understood in detail via \emph{Harmonic Analysis} (\HA) (e.g., \cite{Bk:PSE:SSA:1997:Arrillaga}).
This two-part paper focuses on the formulation and solution of the \emph{Harmonic Power-Flow} (\HPF) problem in three-phase power grids with \CIDER[s].
The assessment of harmonic stability and the design of robust controllers, which are inherently related to the solvability of the \HPF problem%
\footnote{%
    By definition, a system is unstable if the equations describing its behaviour have no equilibrium points \cite{Rep:PSE:SA:2018:Canizares,Rep:PSE:SA:2004:Kundur}.
    For example, unsolvability of the power-flow equations implies that the power system cannot reach a steady state.
},
will be the subject of future work.



\HA can be performed using transient, steady-state, or hybrid methods.
Usually, transient methods work with time-domain models, steady-state methods with frequency-domain models, and hybrid methods with a combination of both.
Time-domain analysis is accurate but computationally intensive, which may not be practical for large systems.
Frequency-domain analysis can be computationally more efficient, but the reliability of the results depends on the accuracy of the models \cite{Jrn:PSE:SSA:1987:Arrillaga} (i.e., in view of approximations such as linearizations).
In this respect, ensuring that a model is accurate, computationally efficient, and generally applicable at the same time is a tough challenge.
According to experience, the last point receives the least priority.
As a result, many of the existing frequency-domain models are only valid for specific devices and controllers, or they neglect the coupling between harmonics.
In order to overcome these limitations, a novel modelling framework for three-phase power grids with \CIDER[s] is developed in this two-part paper.
More precisely, the grid model is based on polyphase circuit theory, and the \CIDER model on \emph{Linear Time-Periodic} (\LTP) systems theory.
Notably, the \CIDER model is modular w.r.t. resource components (i.e., power hardware, control software, and reference calculation), generic w.r.t. control laws (i.e., grid-forming or grid-following behaviour), and accurate w.r.t. the generation and propagation of harmonics (in particular: coupling between harmonics).



The main contributions of this two-part paper are as follows:
\begin{itemize}
    \item
        A generic, modular, and accurate modelling framework for three-phase power grids with \CIDER[s] is developed.
        Notably, the coupling between harmonics is considered.
    \item
        Based on this modelling framework, a formulation of the \HPF problem which can be solved by a single-iterative algorithm via the Newton-Raphson is proposed.
    \item
        A detailed validation of the \HPF method is performed on individual resources and on the \CIGRE low-voltage benchmark microgrid.
        Moreover, the scalability and computational intensity of the method are investigated.
\end{itemize}
\noindent
This paper is divided into two parts:
Part~I focuses on the development of the \HPF method, and Part~II on its validation.%
The remainder of this part is organized as follows.
\cref{sec:review} gives an overview of the state-of-the-art.
\cref{sec:grid,sec:CIDER} present the models of the grid and \CIDER[s], respectively.
\cref{sec:HPF} explains how the \HPF problem is formulated and solved.
\cref{sec:conclusion} draws some first conclusions.

\section{Literature Review}
\label{sec:review}

For the convenience of the reader, the literature review is divided w.r.t. transient, steady-state, and hybrid methods.
Due to space limitations, only selected works are discussed in this paper.
More complete reviews can be found in \cite{Jrn:PSE:SSA:1998:Smith,Jrn:PSE:SSA:2003:Herraiz,Jrn:PSE:SSA:2013:IEEE,Jrn:PSE:SSA:2019:Wang}.


\subsection{Transient Methods}
\label{sec:review:transient}


Transient methods treat the entire system in time domain.
The grid and the connected resources (incl. their controllers) are described by a system of \emph{Differential-Algebraic Equations} (\DAE[s]), which is solved by numerical integration (e.g., using Runge-Kutta methods).
The spectra are then calculated from the obtained waveforms via the \DFT or similar techniques.


Electrical circuits can be studied via nodal analysis (i.e., using nodal equations given by Kirchhoff's current law) or mesh analysis (i.e., using branch equations given by Kirchhoff's voltage law) \cite{Bk:CT:NA:1975:Chua}.
Classical nodal analysis is widely used in power-systems engineering.
It relies on two fundamental hypotheses: i) all voltage and current sources are referenced w.r.t. the ground, and ii) the grid can be represented by a lumped-element model \cite{Bk:CT:NA:1975:Dimo}.
For instance, the \emph{Electromagnetic Transient Program} (\EMTP) \cite{Jrn:PSA:TA:1969:Dommel} employs nodal analysis.
If the circuit contains ungrounded voltage or current sources, the associated branch equations need to be considered, too.
For example, the \emph{Simulation Program with Integrated Circuit Emphasis} (\SPICE) \cite{Rep:CT:NA:1973:Nagel} utilizes this method, which is called \emph{Modified Nodal Analysis} (\MNA) \cite{Jrn:CT:NA:1975:Ho}. 
Some electrical components (e.g., switches or controlled sources) cannot be described by nodal and branch equations alone, so additional equations have to be introduced.
This universal approach, which is known as \emph{Modified Augmented Nodal Analysis} (\MANA), is notably implemented in \EMTP-RV \cite{Jrn:PSA:TA:2007:Mahseredjian}.

Since each component can be represented by a precise time-domain model, transient methods can yield extremely accurate results.
However, this accuracy comes at the cost of computational intensity, which hinders the analysis of large-scale power systems.
Therefore, recent works have looked into the development of models which are both accurate and computationally efficient.
In \cite{Jrn:PSA:TA:2017:Gu}, the converter models are split into slow/nonlinear and fast/linear states, and model-order reduction is performed on the latter.
In \cite{Jrn:PSA:TA:2019:Todeschini}, an average model of a converter is improved with a switching emulator.


\subsection{Steady-State Methods}
\label{sec:review:steady}


If only the steady-state solution (i.e., the periodic waveforms that remain after all transients have died out) are of interest, the analysis can be performed directly in the frequency domain.
Namely, the \DAE[s] from the time domain can be restated as algebraic equations in the frequency domain using the Fourier transform.
The unknowns of these so-called \HPF equations are the harmonic phasors\footnote{Note that the spectrum of a periodic waveform is nonzero only at integer multiplies of the fundamental frequency.} of the time-domain variables \cite{Jrn:PSE:SSA:1995:Arrillaga}.
In specific cases, the harmonic voltages and currents can be approximated by known linear functions of the respective fundamental tones alone.
Under these circumstances, one can first calculate the fundamental voltages and currents in a standard power-flow study, and then infer the harmonic ones (i.e., via the said functions).
This technique is called \emph{Direct Harmonic Analysis} (\DHA) \cite{Bk:PSE:SSA:1997:Arrillaga}.
In general, the \HPF equations are nonlinear, and hence have to be solved iteratively (e.g., using the Newton-Raphson method).
This approach is called \emph{Iterative Harmonic Analysis} (\IHA) \cite{Bk:PSE:SSA:1997:Arrillaga} or \HPF study \cite{Jrn:PSE:SSA:2003:Herraiz}.
\IHA is comparable to transient analysis in terms of accuracy, but its computational cost is substantially lower \cite{Jrn:PSE:SSA:1987:Arrillaga}.
Moreover, this method can be generalized to include interharmonics \cite{Jrn:PSE:SSA:2004:Arrillaga}.


Various works have studied how standard approaches for power-flow analysis can be extended to \HPF study.
In \cite{Jrn:PSE:SSA:1982:Xia}, the nodal equations are solved in the frequency domain using the Newton-Raphson method.
In \cite{Jrn:PSE:SSA:1991:Xu:1}, the branch equations are used instead.
Other researchers use double-iterate methods for \IHA.
In \cite{Jrn:PSE:SSA:1993:Valcarel}, a standard power-flow study is performed at the master-level, and then refined through \IHA at the slave-level.
Notably, these subproblems are formulated in different reference frames, namely phase coordinates \cite{Jrn:PSE:SSA:1968:Laughton} and symmetrical components \cite{Jrn:PSE:SSA:1918:Fortescue}, respectively.
In \cite{Jrn:PSE:SSA:1991:Arrillaga}, the master instead performs a sophisticated \AC/\DC power-flow study.

In \cite{Jrn:PSE:SSA:2020:Chen}, \LTI state-space models are employed for the detection of the fundamental and harmonic currents.
Some works use \LTP systems theory \cite{Ths:CSE:LTP:1991:Wereley}, a generalization of \LTI systems theory, for the analysis of power electronic converters \cite{J:PSA:SSA:2003:Rico,Jrn:PSE:SSA:2017:Kwon,Jrn:PSE:SSA:2019:Wang}.
The \LTP state-space models are first developed in the time domain, and then described in the frequency domain by Toeplitz matrices composed of Fourier coefficients.


Traditionally, engineers and researchers work with models whose structure and parameters are completely known (i.e., white-box models).
In this case, one can perform \HA using analytical methods (e.g., \cite{Jrn:PSE:SSA:2017:Sun}).
However, modern power systems are so complex that both the structure and the parameters of the underlying models are only partially known or even unknown (i.e., grey- or black-box models).
Therefore, data-driven methods, which allow to cope with such lack of information, have recently gained attention.
For instance, one can train an artificial neural network to learn the harmonic model of a \CIDER, such as a photovoltaic generator \cite{Jrn:PSE:SSA:2018:Abbood} or an electric-vehicle charging station \cite{Cnf:PSE:SSA:2019:Liu}.
In \cite{Jrn:PSE:SSA:2020:Nduka}, a recursive least-squares estimator is employed for data-driven \HPF studies.


\subsection{Hybrid Methods}
\label{sec:review:hybrid}

In presence of elements with strongly nonlinear behaviour, \IHA can suffer from convergence problems \cite{Jrn:PSE:SSA:1998:Smith}.
In this case, one can treat the strongly nonlinear resources in time domain and the weakly nonlinear ones in frequency domain (e.g., \cite{Jrn:PSE:HA:1995:Semlyen,Jrn:PSE:HA:2007:Wiechowski}).
Usually, only a handful of resources are analyzed in time domain in order to keep the computational intensity low.
In \cite{Jrn:TSG:TA:2019:Peng}, a dynamic phasor model is used for the simulation and analysis of harmonics in microgrids.


\subsection{Motivation for Further Work in the Field}
\label{sec:review:motivation}


As explained in \cref{sec:intro} and discussed in detail in \cite{Rep:PSE:SA:2018:Canizares}, \ADN[s] are particularly vulnerable to harmonic instability due to the prevalence of \CIDER[s].
Naturally, good understanding of the generation and propagation of harmonics is fundamental for the design of controllers which are robust against harmonic instability.
This requires a suitable method for the formulation and solution of the \HPF equations%
\footnote{%
    Indeed, the system stability is inherently related to the solvability of the system equations (e.g., \cite{Rep:PSE:SA:2004:Kundur,Rep:PSE:SA:2018:Canizares}), in this case the \HPF equations.
}.
Such a method has to be computationally efficient, and the underlying models need to be \emph{generic} (i.e., w.r.t. grid topology and control laws), \emph{modular} (i.e., w.r.t. the components of resources and grid), and \emph{accurate} (i.e., capture the creation and propagation of harmonics through the resources and the grid).
In terms of computational burden, steady-state methods appear to perform better than transient methods.
As to generality and accuracy, the approaches based on \LTP systems theory \cite{J:PSA:SSA:2003:Rico,Jrn:PSE:SSA:2019:Wang} appear promising, but the underlying system models are not modular.


This paper presents an \HPF method for three-phase power grids with \CIDER[s], which is based on polyphase circuit theory and \LTP systems theory.
In contrast to existing approaches, the \CIDER[s] are explicitly divided into modular blocks (i.e., power hardware, control software, and reference calculation), which are coupled via transforms.
This has major advantages.
Firstly, grid-forming and grid-following \CIDER[s] can be represented by the same generic structure.
Secondly, the blocks can be described in different reference frames if needed.
Moreover, this inherently accounts for the propagation of harmonics due to coordinate transformations (e.g., the Park/Clarke transform).
Thirdly, as nonlinear behaviour is confined to one block (i.e., the reference calculation), the numerical analysis is facilitated.
The \HPF problem is defined by hybrid nodal equations of the grid and the closed-loop transfer functions of the \CIDER[s], whose mismatches must be zero in equilibrium.
The resulting system of nonlinear equations is solved numerically by means of the Newton-Raphson method.
Since only the reference calculations of the \CIDER[s] are nonlinear, a single-iterative algorithm is sufficient -- as opposed to the double-iterative algorithms used in many existing approaches (e.g., \cite{Jrn:PSE:SSA:1993:Valcarel,Jrn:PSE:SSA:1991:Arrillaga}).

\section{Model of the Electrical Grid}
\label{sec:grid}

In this section, some fundamental concepts of circuit theory, which the authors of this paper have discussed in \cite{Jrn:CT:NM:2019:Kettner}, are recalled and generalized for the purpose of \HPF analysis of three-phase systems\footnote{Note that \cite{Jrn:CT:NM:2019:Kettner} discusses the modelling of polyphase systems which are in sinusoidal steady state.}.
\cref{sec:grid:circuit} discusses the lumped-element model of the grid, and \cref{sec:grid:parameters} the compound admittance and hybrid matrices which describe the nodal equations.


\subsection{Lumped-Element Model}
\label{sec:grid:circuit}

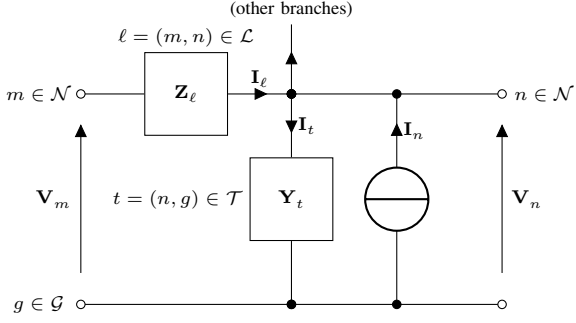
\begin{figure}[t]
	\centering
	{
\tikzstyle{block}=[rectangle, draw=black,minimum size=1.1cm]

\begin{circuitikz}
	
	\scriptsize
	
	
	\def\x{2.8}
	\def\y{2.8}

	
	
	\coordinate (NL) at (-\x,0);	
	\coordinate (NC) at (0,0);
	\coordinate (NR) at (\x,0);
	
	\coordinate (PL) at (-\x,\y);
	\coordinate (PC) at (0,\y);
	\coordinate (PR) at (\x,\y);
	
	
	
	\node[block] (B) at ($0.5*(PL)+0.5*(PC)$) {$\Z_{\ell}$};
	
	\draw (PL)
		to[short,o-] (B.west)
		to[open] (B.east)
		to[short,-*,i=${\I_{\ell}}$] (PC);
	\draw (NL) to[short,o-] (NC);
	
	\node at ($(B)+0.27*(0,\y)$) {$\ell=(m,n)\in\branches$};
	
	\node at ($(PL)-0.2*(\x,0)$) {$m\in\nodes$};
	\node at ($(NL)-0.2*(\x,0)$) {$g\in\ground$};
	
	
	
	\node[block] (S) at ($0.5*(NC)+0.5*(PC)$) {$\Y_{t}$};
	
	\draw (PC)
		to[short,*-,i=${\I_{t}}$] (S.north)
		to[open] (S.south)
		to[short,-*] (NC);
	
	\node at ($(S)-0.54*(\x,0)$) {$t=(n,g)\in\shunts$};
	
	
	
	\draw (NC)
	    to[short,*-] (NR)
	    to[open,o-o,v=${\V_{n}}$] (PR)
	    to[short,-*] (PC);
	    
	\draw ($0.5*(NC)+0.5*(NR)$) to[current source,*-*,i_=${\I_{n}}$] ($0.5*(PC)+0.5*(PR)$);
	\draw (NL) to[open,v^=${\V_{m}}$] (PL);
	
	\node at ($(PR)+0.2*(\x,0)$) {$n\in\nodes$};
	
	
	
	
	\node (BO) at ($(PC)+0.4*(0,\y)$) {(other branches)};
	
	\draw[short,*-,i=$~$] (PC) to (BO.south);
	
\end{circuitikz}

}
	\caption
	{%
	    The grid is represented by branch elements with compound impedance $\Z_{\ell}$ and shunt elements with compound admittance $\Y_{t}$.
	    $\I_{\ell}$ and $\I_{t}$ are the current flows through the branch and shunt elements, respectively.
	    $\V_{n}$ and $\I_{n}$ are the phase-to-ground voltages and injected currents, respectively.
	}
	\label{fig:grid}
\end{figure}

Consider a generic three-phase grid (i.e., radial or meshed, including transposed or non-transposed lines%
\footnote{%
    A line is \emph{transposed} if the positions of its phase conductors are repeatedly swapped long the course of the line, thus guaranteeing by construction that its compound electrical parameters are circulant \cite{Bk:PSE:SSA:1990:Arrillaga}.
},
with balanced or unbalanced nodal injections%
\footnote{%
    The nodal injections or absorptions of a three-phase system are \emph{balanced} if they consist of positive-sequence components only (i.e., the negative- and homopolar-sequence components are null) \cite{Jrn:PSE:SSA:1918:Fortescue}.
}),
which is equipped with a neutral conductor.
Suppose that the neutral conductor is grounded by an effective earthing system, which ensures that the neutral-to-ground voltages are negligible%
\footnote
{%
	Typically, effective earthing systems serve their purpose up to frequencies of a few kilohertz.
	Therefore, this hypothesis is reasonable for \HPF studies, which typically consider harmonics up to order 20-25 (i.e., 1.0-1.5\,kHz).
}.
That is, the phase-to-neutral voltages are equivalent to phase-to-ground voltages, and fully describe the grid state.
Further, assume that the grid can be represented by a set of lumped-element models which are linear and passive (i.e., they contain no active elements like voltage or current sources).
Let $g\in\ground\coloneqq\{0\}$ be the ground and $n\in\nodes$ the nodes, each of which comprises the full set of phase terminals $p\in\phases\coloneqq\{\phsA,\phsB,\phsC\}$.
The lumped elements are divided into branch elements $\ell\in\branches\subseteq\nodes\times\nodes$ and shunt elements $t\in\shunts=\nodes\times\ground$ as illustrated in \cref{fig:grid}.

In line with these considerations, the following hypotheses are made w.r.t. the properties of the grid model:
\begin{Hypothesis}\label{hyp:grid:model}
    Since the lumped elements of the grid model are linear and passive, its circuit equations can be formulated independently at each frequency $f$ using either impedance or admittance parameters.
    Each branch element $\ell$ is described by an impedance equation
    \begin{equation}
        \V_{m}(f)-\V_{n}(f)=\Z_{\ell}(f)\I_{\ell}(f),\quad
    	\forall\ell=(m,n)\in\branches
        \label{eq:branches:law}
    \end{equation}
    where $\Z_{\ell}\in\CompNum^{3\times3}$ is the compound impedance of the branch element, $\I_{\ell}\in\CompNum^{3\times1}$ is the current flowing through it, and $\V_{m},\V_{n}\in\CompNum^{3\times1}$ are the phase-to-ground voltages at its start and end node, respectively.
    Each shunt element $t$ is described by an admittance equation
    \begin{equation}
        \I_{t}(f)=\Y_{t}(f)\V_{n}(f),\quad
    	\forall t=(n,g)\in\shunts
        \label{eq:shunts:law}
    \end{equation}
    where $\Y_{t}\in\CompNum^{3\times3}$ is the compound admittance of the shunt element, and $\I_{t}\in\CompNum^{3\times1}$ the current flowing through it.
\end{Hypothesis}
\noindent
In general, \eqref{eq:branches:law}--\eqref{eq:shunts:law} do not correspond to an \LTI system (i.e., linearity w.r.t. frequency does not guarantee time-invariance).
More precisely, the grid model is \LTI \emph{if and only if} it consists of frequency-independent resistor (or conductor), inductor, and capacitor elements%
\footnote
{%
	For instance, the grid model is \LTI if $\Z_{\ell}(f)=\mathbf{R}_{\ell}+j 2\pi f\mathbf{L}_{\ell}$ ($\forall\ell\in\branches$) and $\Y_{t}(f)=\mathbf{G}_{t}+j 2\pi f\mathbf{C}_{t}$ ($\forall t\in\shunts$) for constant $\mathbf{R}_{\ell}$, $\mathbf{L}_{\ell}$, $\mathbf{G}_{t}$, and $\mathbf{C}_{t}$.
}
(e.g., \cite{Bk:CT:NM:2016:Borghetti}).
Yet, the generic form of \eqref{eq:branches:law}--\eqref{eq:shunts:law} does allow to treat frequency-dependent parameters if needed%
\footnote{%
    E.g., $\Z_{\ell}(f)=\mathbf{R}_{\ell}(f)+j 2\pi f\mathbf{L}_{\ell}(f)$ and $\Y_{t}(f)=\mathbf{G}_{t}(f)+j 2\pi f\mathbf{C}_{t}(f)$.
}.%

Moreover, note that \eqref{eq:branches:law}--\eqref{eq:shunts:law} describe the behaviour of the grid in phase coordinates.
It is important to note that these equations are valid irrespective of any asymmetries in the system w.r.t. the grid components (e.g., due to non-transposition of lines) or the nodal injections/absorptions (e.g., due to unbalances of generation/load) \cite{Jrn:PSE:SSA:1968:Laughton}.
Therefore, these phase-domain equations are particularly suitable for power distribution systems (i.e., unlike sequence-domain equations), where such asymmetries are common.


Note that the term ``compound electrical parameters'' refers to polyphase impedance or admittance matrices, which take into account the electromagnetic coupling between different phases (e.g., \cite{Bk:PSE:SSA:1990:Arrillaga}).
These compound electrical parameters are assumed to have the following properties:
\begin{Hypothesis}\label{hyp:grid:parameters}
	The compound branch impedance matrices $\Z_{\ell}$ are symmetric, invertible, and lossy at all frequencies:
	\begin{equation}
		\Z_{\ell}(f):
		\quad
		\left[~
		\begin{aligned}
			\Z_{\ell}(f)		&=			(\Z_{\ell}(f))^{T}	\\
			\exists\Y_{\ell}(f) &=			(\Z_{\ell}(f))^{-1}	\\
			\Real{\Z_{\ell}(f)}	&\succeq	0
		\end{aligned}
		\right.
		\label{eq:branches:parameters}
	\end{equation}
	The compound shunt admittance matrices $\Y_{t}$ are symmetric, invertible, and lossy at all frequencies if they are nonzero:
	\begin{equation}
		\text{if}~\Y_{t}(f)\neq\mathbf{0}:
		\quad
		\left[~
		\begin{aligned}
			\Y_{t}(f)		    &=          (\Y_{t}(f))^{T}\\
			\exists\Z_{t}(f)    &=          (\Y_{t}(f))^{-1}\\
			\Real{\Y_{t}(f)}    &\succeq	0
		\end{aligned}
		\right.
		\label{eq:shunts:parameters}
	\end{equation}
\end{Hypothesis}
\noindent
These properties follow from fundamental laws of physics (e.g., Maxwell's equations), and hold for a broad variety of grid components.
For instance, the electrical parameters of lines, conventional transformers, and series or shunt compensators satisfy these properties.
Notably, the symmetry property holds for both transposed and non-transposed lines%
\footnote{%
    The compound electrical parameters of a transposed line are \emph{symmetric} and \emph{circulant} (e.g., \cite{Bk:PSE:SSA:1990:Arrillaga}).
    The symmetry property is given by physics, whereas the circulancy property is enforced through construction (i.e., the transposition of the conductors).
    The compound electrical parameters of non-transposed lines are only symmetric.
}.
Only a few types of grid components, such as phase-shifting transformers, exhibit different characteristics (i.e., the symmetry property does not hold).

Observe that \cref{hyp:grid:parameters} refers to the \emph{exact} parameters of the grid components.
In case the exact parameters are not known, they have to be inferred from measurements via system identification (e.g., least-squares regression).
The obtained \emph{estimated} parameters may violate the properties in \cref{hyp:grid:parameters}, unless corresponding constraints are imposed on the solution of the system-identification problem \cite{Jrn:CT:NM:2001:Gustavsen}.


\subsection{Compound Admittance and Hybrid Matrices}
\label{sec:grid:parameters}


The \emph{branch graph} $\branchgraph\coloneqq(\nodes,\branches)$ specifies the grid topology.
Its \emph{three-phase} incidence matrix $\mathbf{A}_{\branchgraph}$ is defined as
\begin{equation}
	\mathbf{A}_{\branchgraph}:~
	\left(\mathbf{A}_{\branchgraph}\right)_{\mathit{kn}}
	\coloneqq	\left\{
				\begin{array}{cl}
					+\diag(\mathbf{1}_{3})	&\text{if $\ell_{k}=(n,\cdot)$}\\
					-\diag(\mathbf{1}_{3})	&\text{if $\ell_{k}=(\cdot,n)$}\\
					\mathbf{0}_{3\times3}	&\text{otherwise}
				\end{array}
				\right.
\end{equation}
where $\diag(\mathbf{1}_{3})$ and $\mathbf{0}_{3\times3}$ are the identity and null matrix, respectively, of size $3\times3$.
Accordingly, $\mathbf{A}_{\branchgraph}\in\RealNum^{3\Card{\branches}\times3\Card{\nodes}}$.%
The \emph{primitive compound admittance matrices} $\Y_{\branches}$ and $\Y_{\shunts}$ associated with the branches and shunts, respectively, are defined as (e.g., see \cite{Jrn:CT:NM:1982:Alvarado})
\begin{alignat}{2}
    \Y_{\branches}(f)
    &\coloneqq          \diag_{\ell\in\branches}(\Y_{\ell}(f))&
    &~\in~   \CompNum^{3\Card{\branches}\times3\Card{\branches}}
    \label{eq:branches:primitive}\\
    \Y_{\shunts}(f)
    &\coloneqq          \diag_{t\in\shunts}(\Y_{t}(f))&
    &~\in~   \CompNum^{3\Card{\nodes}\times3\Card{\nodes}}
    \label{eq:shunts:primitive}
\end{alignat}

Let $\V$ and $\I$ be the vectors of all phase-to-ground voltages and nodal injected currents, respectively:
\begin{alignat}{2}
    \V(f)
    &\coloneqq          \col_{n\in\nodes}(\V_{n}(f))&
    &~\in~   \CompNum^{3\Card{\nodes}\times1}
    \label{eq:nodes:voltages}\\
	\I(f)
	&\coloneqq          \col_{n\in\nodes}(\I_{n}(f))&
	&~\in~   \CompNum^{3\Card{\nodes}\times1}
	\label{eq:nodes:currents}
\end{alignat}
The \emph{compound nodal admittance matrix} $\Y\in\CompNum^{3\Card{\nodes}\times3\Card{\nodes}}$, which links $\I$ to $\V$, is calculated as follows:
\begin{equation}
    \I(f)=\Y(f)\V(f),~
    \Y(f)=\mathbf{A}^{T}_{\branchgraph}\Y_{\branches}(f)\mathbf{A}_{\branchgraph} + \Y_{\shunts}(f)
    \label{eq:nodes:admittance}
\end{equation}



As proven in \cite{Jrn:CT:NM:2019:Kettner}, the following Lemma holds:
\begin{Lemma}\label{lm:grid:hybrid}
	Suppose that \cref{hyp:grid:model,hyp:grid:parameters} hold, the branch graph $\branchgraph$ is weakly connected, and the compound branch impedances $\Z_{\ell}(f)$ are strictly lossy (i.e., $\Real{\Z_{\ell}(f)} \succ 0 ~\forall\ell\in\branches$).
	Further, partition the nodes $\nodes$ into two disjoint sets $\formers$ and $\followers$
	\begin{equation}
	    \nodes=\formers\cup\followers,~\formers\cap\followers=\emptyset
	\end{equation}
	(note that this implies $\Card{\nodes}=\Card{\formers}+\Card{\followers}$).
	Then, the associated block form of the nodal admittance equations \eqref{eq:nodes:admittance}
    \begin{equation}
    	\begin{bmatrix}
    	    \I_{\formers}(f)\\
    	    \I_{\followers}(f)
    	\end{bmatrix}
    	=	\begin{bmatrix}
    			\Y_{\formers\times\formers}(f)      &\Y_{\formers\times\followers}(f)\\
    			\Y_{\followers\times\formers}(f)    &\Y_{\followers\times\followers}(f)\\
    		\end{bmatrix}
    		\begin{bmatrix}
    			\V_{\formers}(f)\\
    			\V_{\followers}(f)
    		\end{bmatrix}
    \end{equation}
	can be reformulated into the nodal hybrid equations
	\begin{equation}
    	    \begin{bmatrix}
    	        \V_{\formers}(f)\\
    	        \I_{\followers}(f)
    	    \end{bmatrix}
        =   \begin{bmatrix}
    	            \HG_{\formers\times\formers}(f)
                &   \HG_{\formers\times\followers}(f)\\
                    \HG_{\followers\times\formers}(f)
                &   \HG_{\followers\times\followers}(f)
    	    \end{bmatrix}
    	    \begin{bmatrix}
    	        \I_{\formers}(f)\\
    	        \V_{\followers}(f)
    	    \end{bmatrix}
        \label{eq:nodes:hybrid}
	\end{equation}
	The blocks of the compound hybrid matrix $\HG$ are given by
	\begin{alignat}{2}
	    \HG_{\formers\times\formers}(f)
        &=                  \phantom{-}\Y_{\formers\times\formers}^{-1}(f)&
        &~\in~   \CompNum^{3\Card{\formers}\times3\Card{\formers}}
        \\
        \HG_{\formers\times\followers}(f)
        &=                  -\Y_{\formers\times\formers}^{-1}(f)\Y_{\formers\times\followers}(f)&
        &~\in~   \CompNum^{3\Card{\formers}\times3\Card{\followers}}
        \\
        \HG_{\followers\times\formers}(f)
        &=                  \phantom{-}\Y_{\followers\times\formers}(f)\Y_{\formers\times\formers}^{-1}(f)&
        &~\in~   \CompNum^{3\Card{\followers}\times3\Card{\formers}}
        \\
        \HG_{\followers\times\followers}(f)
        &=                  \phantom{-}\Y(f)/\Y_{\followers\times\followers}(f)&
        &~\in~   \CompNum^{3\Card{\followers}\times3\Card{\followers}}
	\end{alignat}
	where $\Y/\Y_{\followers\times\followers}$ is the \emph{Schur complement} of $\Y$ w.r.t. $\Y_{\followers\times\followers}$.
\end{Lemma}
\noindent
In the \HPF study, $\formers$ and $\followers$ are the nodes with grid-forming and grid-following \CIDER[s], respectively (see \cref{sec:HPF}).

Observe that \eqref{eq:nodes:hybrid} holds for any arbitrary frequency $f$ (i.e., as stated in \cref{hyp:grid:model} and \ref{hyp:grid:parameters}).
Now, consider the special case of \emph{harmonic frequencies} $f_{h}$, which are defined by the \emph{harmonic orders} $h\in\harmonics$ w.r.t. a given \emph{fundamental frequency} $f_{1}$:

\begin{equation}
    f_{h} \coloneqq h \cdot f_{1}, ~ h\in\harmonics\subset\mathbb{Z}
    \label{eq:harmonics}
\end{equation}
Due to the assumed linearity of the grid components (recall \cref{hyp:grid:model}), the hybrid nodal equations \eqref{eq:nodes:hybrid} can be formulated separately at each of the harmonic frequencies $f_{h}$.
Combining the equations for the individual harmonic frequencies yields a system of equations for the entire harmonic spectrum
\begin{equation}
    \begin{bmatrix}
	        \hat{\V}_{\formers}\\
	        \hat{\I}_{\followers}
	    \end{bmatrix}
    =   \begin{bmatrix}
	            \hat{\HG}_{\formers\times\formers}
            &   \hat{\HG}_{\formers\times\followers}\\
                \hat{\HG}_{\followers\times\formers}
            &   \hat{\HG}_{\followers\times\followers}
	    \end{bmatrix}
	    \begin{bmatrix}
	        \hat{\I}_{\formers}\\
	        \hat{\V}_{\followers}
	    \end{bmatrix}
	   \label{eq:nodes:hybrid:harmonics}
\end{equation}
where
\begin{alignat}{2}
    \hat{\V}_{\formers}
    &\coloneqq          \col_{h\in\harmonics}(\V_{\formers}(f_{h}))&
    &~\in~   \CompNum^{3\Card{\harmonics}\Card{\formers}\times1}
    \\
    \hat{\HG}_{\formers\times\formers}
    &\coloneqq          \diag_{h\in\harmonics}(\HG_{\formers\times\formers}(f_{h}))&
    &~\in~   \CompNum^{3\Card{\harmonics}\Card{\formers}\times3\Card{\harmonics}\Card{\formers}}
\end{alignat}
The remaining blocks of $\hat{\V}$, $\hat{\I}$, and $\hat{\HG}$ are defined analogously.

\section{Model of Converter-Interfaced Distributed Energy Resources}
\label{sec:CIDER}

In this section, the \CIDER model is proposed.
First, the time-domain state-space model is presented in \cref{sec:CIDER:time}.
Then, the harmonic-domain state-space model is derived using \LTP systems theory in \cref{sec:CIDER:harmonic}.


\subsection{Time-Domain State-Space Model}
\label{sec:CIDER:time}

Depending on the operating mode, a \CIDER[s] is classified as either \emph{grid-forming} or \emph{grid-following} (e.g., \cite{Rep:PSE:SA:2018:Canizares}):
\begin{Definition}\label{def:CIDER:forming}
	A grid-forming \CIDER controls the magnitude and frequency of the grid voltage at its point of connection.
\end{Definition}
\begin{Definition}\label{def:CIDER:following}
	A grid-following \CIDER controls the injected current with a specific phase displacement w.r.t. the fundamental component of the grid voltage at its point of connection.
	This requires a grid-synchronization mechanism which provides knowledge of the fundamental-frequency phasor of the grid voltage (e.g., a \PLL).
\end{Definition}
\noindent
That is, a grid-forming \CIDER behaves like a controlled voltage source with finite output impedance, and a grid-following \CIDER like a controlled current source.
As will be shown shortly, either behaviour can be represented by a transfer function which characterizes the creation and propagation of harmonics by the respective type of resource.



\begin{figure}[t]
    \centering
    {

\tikzstyle{block-big}=[rectangle,draw=black,minimum width=1.4cm,minimum height=1.0cm,inner sep=0pt]
\tikzstyle{conversion}=[rectangle,draw=black,minimum width=0.9cm,minimum height=0.6cm,inner sep=0pt]
\tikzstyle{transformation}=[rectangle,draw=black,minimum width=0.7cm,minimum height=0.7cm,inner sep=0pt]
\tikzstyle{controller}=[rectangle,draw=black,minimum width=1.0cm,minimum height=1.0cm,inner sep=0pt]

\tikzstyle{dot}=[circle,draw=black,fill=black,inner sep=1pt]
\tikzstyle{signal}=[-latex]

\scriptsize

\begin{tikzpicture}
    \def\dx{1.0}
    \def\dy{1.0}
    
    
    
    \node[block-big] (A) at (0,0)
    {%
    \begin{tabular}{c}
        Actuator\\
        $\act$
    \end{tabular}
    };
    
    \node[block-big] (F) at ($(A)+(2.2*\dx,0)$)
    {%
    \begin{tabular}{c}
        Filter\\
        $\fltr_{\lambda}$
    \end{tabular}
    };
    \node[dot] (FO) at ($(F)-(0,\dy)$) {};
    \draw[signal] (A.east) to (F.west);
    
    \node[dot] (PCI) at ($(F)+(2.2*\dx,0)$) {};
    \coordinate (PCO) at ($(PCI)-(0,\dy)$);
    
    \node[transformation] (TGI) at ($(PCI)+(\dx,0)$) {$\tau_{\pwr|\gamma}$};
    \node[transformation] (TGO) at ($(PCO)+(\dx,0)$) {$\tau_{\gamma|\pwr}$};
    
    \node (GI) at ($(TGI)+(1.2*\dx,0)$) {$\mathbf{w}_{\grid}$};
    \node (GO) at ($(TGO)+(1.2*\dx,0)$) {$\mathbf{y}_{\grid}$};
    \node (G) at ($0.5*(GI)+0.5*(GO)$) {Grid $\grid$};
    
    \draw[signal] (GI.west) to (TGI.east);
    \draw[signal] (TGO.east) to (GO.west);
    
    \draw (TGI.west) to (PCI.east);
    \draw[signal] (PCI.east) to node[midway,above]{$\mathbf{w}_{\pwr}$} (F.east);
    \draw[signal] (FO.east) to node[midway,above]{$\mathbf{y}_{\pwr,\Lambda}$} (TGO.west);
    
    
    
    \node[conversion] (FAI) at ($(A)-(0,2.8*\dy)$) {\LPF};
    \draw[signal] (FAI.north) to node[near end,right]{$\mathbf{u}_{\pwr}$} (A.south);
    
    \node[conversion] (FAAI) at ($(F)-2.8*(0,\dy)$) {\LPF};
    \node[conversion] (FAAO) at ($(PCI)-2.8*(0,\dy)$) {\LPF};
    \draw (F.south) to (FO.north);
    \draw[signal] (FO.south) to node[near start,right]{$\mathbf{y}_{\pwr,\lambda}$} (FAAI.north);
    \draw[signal] (PCI.south) to (FAAO.north);
    
    \node[conversion] (DAC) at ($(FAI)-(0,\dy)$) {\DAC};
    \draw[signal] (DAC.north) to (FAI.south);
    
    \node[conversion] (ADCI) at ($(FAAI)-(0,\dy)$) {\ADC};
    \node[conversion] (ADCO) at ($(FAAO)-(0,\dy)$) {\ADC};
    \draw[signal] (FAAI.south) to (ADCI.north);
    \draw[signal] (FAAO.south) to (ADCO.north);
    
    \node[transformation] (TBW) at ($(DAC)-(0,\dy)$) {$\tau_{\pwr|\ctrl}$};
    
    \node[transformation] (TFWI) at ($(ADCI)-(0,\dy)$) {$\trafo_{\ctrl|\pwr}$};
    \node[transformation] (TFWO) at ($(ADCO)-(0,\dy)$) {$\tau_{\ctrl|\pwr}$};
    
    \draw[signal] (TBW.north) to (DAC.south);
    \draw[signal] (ADCI.south) to (TFWI.north);
    \draw[signal] (ADCO.south) to (TFWO.north);
    
    
    
    \node[block-big] (K) at ($(TFWI)-1.5*(0,\dy)$)
    {%
    \begin{tabular}{c}
        Controller\\
        $\ctrl_{\lambda}$
    \end{tabular}
    };
    \node[block-big] (R) at ($(TFWO)-1.5*(0,\dy)$)
    {%
    \begin{tabular}{c}
        Reference\\
        $\refr$
    \end{tabular}
    };
    
    \node (SP) at ($(R)+(2.2*\dx,0)$) {$\mathbf{w}_{\spt}$};
    \node at ($(SP)+(0,0.4*\dy)$) {Setpoint $\spt$};
    
    \draw[signal] (TFWI.south) to node[midway,right]{$\mathbf{u}_{\ctrl,\lambda}$} (K.north);
    \draw[signal] (TFWO.south) to node[midway,right]{$\mathbf{w}_{\refr}$} (R.north);
    
    \draw[signal] (SP.west) to (R.east);
    \draw[signal] (R.west) to node[midway,above]{$\mathbf{w}_{\ctrl}$} (K.east);
    \draw[signal] (K.west) to node[midway,above]{$\mathbf{y}_{\ctrl}$} ($(TBW)-1.5*(0,\dy)$) to (TBW.south);
    
    
    
    
    
    \draw[dashdotted] ($(F.north east)+0.15*(\dx,\dy)$)
        to ($(K.south east)+0.15*(\dx,-\dy)$)
        to ($(K.south west)+0.15*(-\dx,-\dy)$)
        to ($(F.north west)+0.15*(-\dx,\dy)$)
        to ($(F.north east)+0.15*(\dx,\dy)$);
    \node at ($(F.north)+0.4*(0,\dy)$) {Cascaded Loops $\lambda\in\{1,\ldots,\Lambda\}$};
    
    
    
    \coordinate (DL) at ($(DAC)-0.8*(\dx,0)$);
    \coordinate (DR) at ($(ADCO)+2.8*(\dx,0)$);
    
    \draw[dotted] (DL) to (DAC.west);
    \draw[dotted] (DAC.east) to (ADCI.west);
    \draw[dotted] (ADCI.east) to (ADCO.west);
    \draw[dotted] (ADCO.east) to (DR);
    
    \node at ($0.85*(DR)+0.15*(DL)+0.4*(0,\dy)$)
    {%
    \begin{tabular}{c}
        Analog Subsystem\\
        (Continuous-Time)
    \end{tabular}
    };
    \node at ($0.85*(DR)+0.15*(DL)-0.4*(0,\dy)$){%
    \begin{tabular}{c}
        Digital Subsystem\\
        (Discrete-Time)
    \end{tabular}
    };
    
    
    
    \coordinate (ML) at ($(FAI)-0.8*(\dx,0)+0.9*(0,\dy)$);
    \coordinate (MR) at ($(FAAO)+2.8*(\dx,0)+0.9*(0,\dy)$);
    
    \draw[dashed] (ML) to (MR);
    
    \node at ($0.85*(MR)+0.15*(ML)+0.2*(0,\dy)$){Power Hardware $\pwr$};
    \node at ($0.85*(MR)+0.15*(ML)-0.2*(0,\dy)$){Control Software $\ctrl$};
    
\end{tikzpicture}

}
    \caption
    {%
        Schematic diagram of a generic \CIDER.
        The power hardware $\pwr$ consists of the actuator $\act$ and cascaded filters $\fltr_{\lambda}$ (for simplicity, one stage is shown only), the outermost of which is connected to the grid $\grid$.
        The control software $\ctrl$ consists of the reference calculation $\refr$ and cascaded controllers $\ctrl_{\lambda}$, which track the setpoint $\spt$.
        The transforms $\trafo$ represent changes of electrical connection or reference frame.
    }
    \label{fig:CIDER:schematic}
\end{figure}
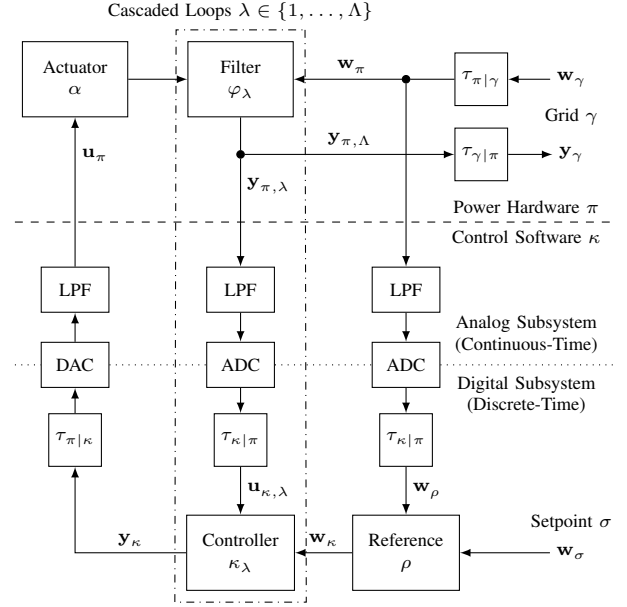

Both types of \CIDER[s] have the same generic structure shown in \cref{fig:CIDER:schematic}: they consist of \emph{power hardware} $\pwr$ and \emph{control software} $\ctrl$.
The power hardware consists of an \emph{actuator} $\act$, for instance a full-wave or half-wave bridge, and a \emph{filter}
$\varphi$, for example an L, LC, LCL or higher-order filter.
The filter consists of filter stages $\varphi_{\lambda}$, whose state variables (i.e., currents through inductors or voltages across capacitors) can be controlled if desired.
Therefore, in general, each filter stage $\varphi_{\lambda}$ in the power hardware can be coupled with a corresponding \emph{controller} stage $\kappa_{\lambda}$ in the control software.
Each pair of filter and controller stage forms a control loop, as illustrated in \cref{fig:CIDER:schematic}. 
However, in practice, it may not be necessary to use a controller stage for each and every filter stage%
\footnote{%
    For instance, it is common practice to control only the current through either the grid-side or the actuator-side inductor of an LCL filter (e.g., \cite{Jrn:PSE:C:2016:Xin}).
}
(i.e., a \CIDER can have fewer controller stages than filter stages).
In such cases, the schematic in \cref{fig:CIDER:schematic} can be simplified accordingly.
Internally, each controller stage may comprise several parallel controllers (e.g., for the mitigation of specific harmonics).

Additionally, the control software contains the \emph{reference calculation} $\refr$, which computes the reference signal for the controller from the \emph{setpoint} $\spt$ (i.e., voltage magnitude and frequency for grid-forming \CIDER[s], active/reactive power for grid-following ones).
These setpoints are provided by system-level controllers (e.g., tertiary controls), which act on a significantly slower timescale than the resource-level controllers (e.g., primary and secondary controls).
That is, they move the equilibrium by changing the setpoints, but have no impact on the harmonics (e.g., \cite{Jrn:PSE:C:2012:Guerrero:1,Jrn:PSE:C:2012:Guerrero:2}).
Hence, the system-level controllers can be neglected for the purpose of steady-state analysis.

Observe that power hardware and control software are connected in a circular fashion: one's outputs are the other's inputs.
The power hardware subsystem is analog and continuous-time, whereas the control software subsystem is digital and discrete-time.
These subsystems are interfaced via \emph{Analog-to-Digital Converters} (\ADC[s]) and \emph{Digital-to-Analog Converters} (\DAC[s]), which are coupled with \emph{Low-Pass Filters} (\LPF[s]) for anti-aliasing and anti-imaging, respectively.
In general, the models of the grid, power hardware, and control software are formulated in different reference frames.
This is represented by the transforms $\trafo_{\ctrl|\pwr}$ and $\trafo_{\pwr|\ctrl}$ in \cref{fig:CIDER:schematic}.
Moreover, the electrical connections of grid and power hardware may be different (e.g., four-wire lines vs. three-leg or four-leg power converters).
This is captured by the transforms $\trafo_{\pwr|\grid}$ and $\trafo_{\grid|\pwr}$ in \cref{fig:CIDER:schematic}.



For the purpose of \HPF study, the system is assumed to be in periodic steady state:
\begin{Hypothesis}\label{hyp:periodic-steady-state}
    There exists a steady state in which all time-variant quantities are periodic with \emph{period} $T$.
    That is, the system behaviour is characterized by the \emph{fundamental frequency} $f_{1}$ and the \emph{harmonic orders} $h\in\harmonics$.
\end{Hypothesis}
\noindent
The existence of a steady-state solution, as well as its location in the solution space, depend on the setpoints imposed by the system-level controllers.
In the periodic state, the components of the \CIDER as shown in \cref{fig:CIDER:schematic} are described by \LTP models.%
    

As previously mentioned, the power hardware is an analog continuous-time system.
It is represented by the \LTP model
\begin{align}
        \dot{\mathbf{x}}_{\pwr}(t)
    &=  \mathbf{A}_{\pwr}(t)\mathbf{x}_{\pwr}(t)
		+   \mathbf{B}_{\pwr}(t)\mathbf{u}_{\pwr}(t)
		+   \mathbf{E}_{\pwr}(t)\mathbf{w}_{\pwr}(t)
    \label{eq:pwr:proc:time}\\
        \mathbf{y}_{\pwr}(t)
	&=      \mathbf{C}_{\pwr}(t)\mathbf{x}_{\pwr}(t)
        +   \mathbf{D}_{\pwr}(t)\mathbf{u}_{\pwr}(t)
		+   \mathbf{F}_{\pwr}(t)\mathbf{w}_{\pwr}(t)
	\label{eq:pwr:meas:time}
\end{align}
$\mathbf{x}_{\pwr}(t)$, $\mathbf{u}_{\pwr}(t)$, $\mathbf{y}_{\pwr}(t)$, and $\mathbf{w}_{\pwr}(t)$ are the \emph{state}, \emph{input}, \emph{output}, and \emph{disturbance vector}, respectively, of the power hardware.
Accordingly, $\mathbf{A}_{\pwr}(t)$, $\mathbf{B}_{\pwr}(t)$, $\mathbf{C}_{\pwr}(t)$, $\mathbf{D}_{\pwr}(t)$, $\mathbf{E}_{\pwr}(t)$, and $\mathbf{F}_{\pwr}(t)$ are the \emph{system}, \emph{input}, \emph{output}, \emph{feed-through}, \emph{input disturbance}, and \emph{output disturbance matrix}, respectively.
The sizes of these vectors and matrices depend on the reference frame in which the power hardware is modelled%
\footnote{%
    If phase coordinates are used, $\mathbf{x}_{\pwr},\mathbf{y}_{\pwr}\in\RealNum^{3\Lambda\times1}$ and $\mathbf{u}_{\pwr},\mathbf{y}_{\pwr}\in\RealNum^{3\times1}$.
    The sizes of the matrices follow from \eqref{eq:pwr:proc:time}--\eqref{eq:pwr:meas:time}.
}.%
The transforms linking the grid and the power hardware are described by
\begin{alignat}{2}
    \tau_{\pwr|\grid}:
    &~
    &   \mathbf{w}_{\pwr}(t)
    &=  \mathbf{T}_{\pwr|\grid}(t)\mathbf{w}_{\grid}(t)
    \label{eq:trafo:grd-to-pwr:time}\\
    \tau_{\grid|\pwr}:
    &~
    &   \mathbf{y}_{\grid}(t)
    &=  \mathbf{T}^{+}_{\grid|\pwr}(t)\mathbf{y}_{\pwr}(t),~
        [\mathbf{T}^{+}_{\grid|\pwr}(t)]_{1,\Lambda} = \mathbf{T}_{\grid|\pwr}(t)
    \label{eq:trafo:pwr-to-grd:time}
\end{alignat}
where $\mathbf{T}_{\pwr|\grid}(t)$ and $\mathbf{T}_{\grid|\pwr}(t)$ are the associated transformation matrices.
Only one column block of $\mathbf{T}^{+}_{\grid|\pwr}(t)$ is nonzero, since $\mathbf{y}_{\grid}(t)$ includes only the block $\mathbf{y}_{\pwr,\Lambda}(t)$ of $\mathbf{y}_{\pwr}(t)$.
This \LTP form is generic: for instance, it allows to represent the behaviour of switching equipment (e.g.,  \cite{Cnf:PSE:SSA:2008:Love,Jrn:PSE:SSA:2019:Wang}).
If this is not required, the \LTP equations become \LTI (i.e., a trivial case of periodic).



The control software is a digital discrete-time system.
In this respect, the following hypothesis is made:
\begin{Hypothesis}\label{hyp:nyquist-shannon}
    The \ADC[s], \DAC[s], and their \LPF[s] (see \cref{fig:CIDER:schematic}) are designed such that an exact reconstruction of the signals is feasible in the frequency band of interest for \HPF studies (i.e., in line with the Nyquist-Shannon sampling theorem).
\end{Hypothesis}
\noindent
That is, the effects of sampling and quantization in the \ADC[s] and reconstruction in the \DAC[s] can be neglected.
Therefore, the control software can be represented by an equivalent continuous-time model.
Analogous to \eqref{eq:pwr:proc:time}--\eqref{eq:pwr:meas:time} of the power hardware, the control software is described by the \LTP system
\begin{align}
        \dot{\mathbf{x}}_{\ctrl}(t)
    &=  \mathbf{A}_{\ctrl}(t)\mathbf{x}_{\ctrl}(t)
		+   \mathbf{B}_{\ctrl}(t)\mathbf{u}_{\ctrl}(t)
		+   \mathbf{E}_{\ctrl}(t)\mathbf{w}_{\ctrl}(t)
    \label{eq:ctrl:proc:time}\\
        \mathbf{y}_{\ctrl}(t)
	&=      \mathbf{C}_{\ctrl}(t)\mathbf{x}_{\ctrl}(t)
        +   \mathbf{D}_{\ctrl}(t)\mathbf{u}_{\ctrl}(t)
		+   \mathbf{F}_{\ctrl}(t)\mathbf{w}_{\ctrl}(t)
	\label{eq:ctrl:meas:time}
\end{align}
The size of the matrices and vectors depends on the frame of reference in which the control software is modelled%
\footnote{%
    In case direct-quadrature components are used, $\mathbf{x}_{\ctrl},\mathbf{y}_{\ctrl}\in\RealNum^{2\Lambda\times1}$ and $\mathbf{u}_{\ctrl},\mathbf{y}_{\ctrl}\in\RealNum^{2\times1}$.
    The sizes of the matrices follow from \eqref{eq:ctrl:proc:time}--\eqref{eq:ctrl:meas:time}.
}.%

As previously mentioned, in general, each filter stage can be coupled with a controller stage.
Accordingly, each control loop $\lambda$ is associated with a corresponding block in $\mathbf{y}_{\pwr}(t)$ and $\mathbf{u}_{\ctrl}(t)$:
\begin{align}
    \mathbf{y}_{\pwr}(t)    &=   \col_{\lambda}\left(\mathbf{y}_{\pwr,\lambda}(t)\right) \\
    \mathbf{u}_{\ctrl}(t)   &=   \col_{\lambda}\left(\mathbf{u}_{\ctrl,\lambda}(t)\right)
\end{align}
If some filter stages are not coupled with a controller stage, the associated blocks can simply be omitted.

Recall from \cref{fig:CIDER:schematic} that the outputs of the power hardware are connected to the inputs of the control software, and vice versa.
Since the influence of the \ADC[s], \DAC[s], and \LPF[s] can be neglected according to \cref{hyp:nyquist-shannon}, only the transforms remain:
\begin{align}
        \mathbf{u}_{\ctrl}(t)
    &=  \mathbf{T}^{+}_{\ctrl|\pwr}(t)\mathbf{y}_{\pwr}(t),~
        \mathbf{T}^{+}_{\ctrl|\pwr}(t) = \diag_{\lambda}\left(\mathbf{T}_{\ctrl|\pwr}(t)\right)
    \label{eq:trafo:pwr-to-ctrl:time}\\
        \mathbf{u}_{\pwr}(t)
    &=  \mathbf{T}_{\pwr|\ctrl}(t)\mathbf{y}_{\ctrl}(t)
    \label{eq:trafo:ctrl-to-pwr:time}\\
        \mathbf{w}_{\refr}(t)
    &=  \mathbf{T}_{\ctrl|\pwr}(t)\mathbf{w}_{\pwr}(t)
    \label{eq:trafo:pwr-to-ref:time}
\end{align}
The Clarke \cite{Jrn:PSE:CA:1951:Duesterhoeft} and Park \cite{Jrn:PSE:CA:1929:Park} transform are notable examples, which are widely used.
In general, the transformation matrices are rectangular%
\footnote
{%
	It is common to model the power hardware in phase ($\phsA\phsB\phsC$) coordinates and the control software in direct-quadrature ($\cmpD\cmpQ$) components, respectively.
	In this case, $\mathbf{T}_{\ctrl|\pwr}=\mathbf{T}_{\cmpD\cmpQ|\phsA\phsB\phsC}\in\mathbb{R}^{2\times3}$ and $\mathbf{T}_{\pwr|\ctrl}=\mathbf{T}_{\phsA\phsB\phsC|\cmpD\cmpQ}\in\mathbb{R}^{3\times2}$.
} (i.e., not necessarily square).

As specified in \cref{fig:CIDER:schematic} and \eqref{eq:pwr:proc:time}--\eqref{eq:pwr:meas:time}, the grid acts both as a disturbance and an output from the point of view of the power hardware.
Whether the phase-to-ground voltage $\mathbf{v}(t)$ or the injected current $\mathbf{i}(t)$ at the point of connection is the disturbance or output, depends on the operating mode of the \CIDER.
According to \cref{def:CIDER:forming,def:CIDER:following}:
\begin{align}
    \mathbf{w}_{\grid}(t)
	&\sim	\left\{
				\begin{array}{cl}
					\mathbf{i}(t)	&\text{if \CIDER is grid-forming}	\\
					\mathbf{v}(t)	&\text{if \CIDER is grid-following}
				\end{array}
				\right.
	\label{eq:grid:disturbance:time}\\
	\mathbf{y}_{\grid}(t)
	&\sim	\left\{
				\begin{array}{cl}
					\mathbf{v}(t)	&\text{if \CIDER is grid-forming}	\\
					\mathbf{i}(t)	&\text{if \CIDER is grid-following}
				\end{array}
				\right.
	\label{eq:grid:output:time}
\end{align}
Similarly, as specified in \cref{fig:CIDER:schematic} and \eqref{eq:ctrl:proc:time}--\eqref{eq:ctrl:meas:time}, the setpoint is a disturbance from the point of view of the control software.
In view of \cref{def:CIDER:forming,def:CIDER:following}:
\begin{align}
    \mathbf{w}_{\spt}(t)
	&\sim	\left\{
				\begin{array}{cl}
					V,f &\text{if \CIDER is grid-forming}	\\
					P,Q &\text{if \CIDER is grid-following}
				\end{array}
				\right.
	\label{eq:setpoint:time}
\end{align}
\noindent
The reference calculation is described by the function $\mathbf{r}(\cdot,\cdot)$
\begin{equation}
	\refr:\quad
	\mathbf{w}_{\ctrl}(t) = \mathbf{r}\left(\mathbf{w}_{\refr}(t),\mathbf{w}_{\spt}(t)\right)
	\label{eq:ref:time}
\end{equation}
It is important to note that $\mathbf{r}(\cdot,\cdot)$ need not be linear.
For grid-following \CIDER[s] (i.e., with $\mathit{PQ}$ control), which compose the majority of resources in a power grid, $\mathbf{r}(\cdot,\cdot)$ is nonlinear.
For grid-forming \CIDER[s] (i.e., with $\mathit{Vf}$ control), which are the minority of resources (typically only one), $\mathbf{r}(\cdot,\cdot)$ is linear.
The fact that only a small part of the \CIDER model (i.e., the reference calculation) may be nonlinear, whereas most of the \CIDER model (i.e., \LTP systems and \LTP transforms) is exactly linear, is crucial for computational efficiency.
This will be discussed in more detail later.


\begin{figure}[t]
    \centering
    {

\tikzstyle{system}=[rectangle, draw=black, minimum width=4.0cm, minimum height=1.0cm, inner sep=0pt]
\tikzstyle{block}=[rectangle, draw=black, minimum size=0.8cm, inner sep=0pt]
\tikzstyle{dot}=[circle, draw=black, fill=black, minimum size=0.1cm, inner sep=0pt]

\tikzstyle{signal}=[-latex]

\scriptsize

\begin{tikzpicture}

	\def\dx{1.0}
	\def\dy{1.0}
	
	
	
	\node[system] (P) at (0,1.2*\dy)
	{%
	$
	\begin{aligned}
		    \dot{\mathbf{x}}_{\pwr}
		&=      \mathbf{A}_{\pwr}\mathbf{x}_{\pwr}
		    +   \mathbf{B}_{\pwr}\mathbf{u}_{\pwr}
		    +   \mathbf{E}_{\pwr}\mathbf{w}_{\pwr}
	    \\
		    \mathbf{y}_{\pwr}
		&=      \mathbf{C}_{\pwr}\mathbf{x}_{\pwr}
		    +   \mathbf{D}_{\pwr}\mathbf{u}_{\pwr}
		    +   \mathbf{F}_{\pwr}\mathbf{w}_{\pwr}
	\end{aligned}
	$
	};
	
	\node at ($(P.north)+(0,0.3*\dy)$) {Power Hardware $\pwr$};
	
	\node[system] (C) at (0,-1.2*\dy)
	{%
	$
	\begin{aligned}
		    \dot{\mathbf{x}}_{\ctrl}
		&=      \mathbf{A}_{\ctrl}\mathbf{x}_{\ctrl}
		    +   \mathbf{B}_{\ctrl}\mathbf{u}_{\ctrl}
		    +   \mathbf{E}_{\ctrl}\mathbf{w}_{\ctrl}
	    \\
		    \mathbf{y}_{\ctrl}
		&=      \mathbf{C}_{\ctrl}\mathbf{x}_{\ctrl}
		    +   \mathbf{D}_{\ctrl}\mathbf{u}_{\ctrl}
		    +   \mathbf{F}_{\ctrl}\mathbf{w}_{\ctrl}
	\end{aligned}
	$
	};
	
	\node at ($(C.north)+(0,0.3*\dy)$) {Control Software $\ctrl$};
	
	\node[block] (TPC) at ($0.5*(P.east)+0.5*(C.east)+(0.4*\dx,0)$) {$\mathbf{T}_{\ctrl|\pwr}$};
	\node[block] (TCP) at ($0.5*(P.west)+0.5*(C.west)-(0.4*\dx,0)$) {$\mathbf{T}_{\pwr|\ctrl}$};
	
	\node[dot] (YP) at ($(P.east)+(0.4*\dx,0)$) {};
	
	\draw[-] (P.east) to (YP.west);
	\draw[signal] (YP.south) to node[right]{$\mathbf{y}_{\pwr,\lambda}$} (TPC.north);
	\draw[signal] (TPC.south)
		to node[right]{$\mathbf{u}_{\ctrl,\lambda}$} ($(C.east)+(0.4*\dx,0.3*\dy)$)
		to ($(C.east)+(0,0.3*\dy)$);
	
	\draw[signal] (C.west)
		to ($(C.west)-(0.4*\dx,0)$)
		to node[left]{$\mathbf{y}_{\ctrl}$} ($(TCP.south)$);
	\draw[signal] (TCP.north)
		to node[left]{$\mathbf{u}_{\pwr}$} ($(P.west)-(0.4*\dx,0.3*\dy)$)
		to ($(P.west)-(0,0.3*\dy)$);
		
	
	\node[block] (TPR) at ($(TPC)+(\dx,0)$) {$\mathbf{T}_{\ctrl|\pwr}$};
	\node[dot] (WP) at ($(TPR)+(0,2.4*\dy)$) {};
	\node[block] (TGP1) at ($(WP)+(\dx,0)$) {$\mathbf{T}_{\pwr|\grid}$};
	\node[block] (TGP2) at ($(TGP1)-(0,1.2*\dy)$) {$\mathbf{T}_{\grid|\pwr}$};
	\node (WG) at ($(TGP1)+(1.2*\dx,0)$) {$\mathbf{w}_{\grid}$};
	\node (YG) at ($(TGP2)+(1.2*\dx,0)$) {$\mathbf{y}_{\grid}$};
	
	\draw[signal] (WG.west) to (TGP1.east);
	\draw[signal] (TGP2.east) to (YG.west);
	\draw[-] (TGP1.west) to node[near end,above]{$\mathbf{w}_{\pwr}$} (WP.east) ;
	\draw[signal] (WP.south) to (TPR.north);
	\draw[signal] (WP.west)
		to ($(P.west)+(-0.4*\dx,1.2*\dy)$)
		to ($(P.west)+(-0.4*\dx,0.3*\dy)$)
		to ($(P.west)+(0,0.3*\dy)$);
	\draw[signal] (YP.east) to node[near start,above]{$\mathbf{y}_{\pwr,\Lambda}$} (TGP2.west);
	
	\node at ($0.5*(WG)+0.5*(YG)$) {Grid $\grid$};
	
	
	
	\node[block] (R) at ($(TPR)-(0,1.5*\dy)$) {$\mathbf{r}(\cdot,\cdot)$};
	\node (SP) at ($(R)+1.7*(\dx,0)$) {$\mathbf{w}_{\spt}$};
	
	\draw[signal] (SP.west) to (R.east);
	\draw[signal] (TPR.south) to node[right]{$\mathbf{w}_{\refr}$} (R.north);
	\draw[signal] (R.west) to node[above]{$\mathbf{w}_{\ctrl}$} ($(C.east)-(0,0.3*\dy)$);
	
	\node at ($(SP)+0.4*(0,\dy)$) {Setpoint $\spt$};
	
\end{tikzpicture}

}
    \caption
    {%
        Block diagram of the proposed generic state-space model of \CIDER[s].
        Note the modularity: power hardware $\pwr$, control software $\ctrl$, and grid $\grid$ are represented by separate blocks, which are interfaced via coordinate transformations.
        The reference calculation $\mathbf{r}(\cdot,\cdot)$ may be either linear (i.e., for $\mathit{Vf}$ control) or nonlinear (i.e., for $\mathit{PQ}$ control).
        The other blocks of the model are exactly linear (i.e., \LTP systems and \LTP transforms).
    }
    \label{fig:CIDER:model}
\end{figure}
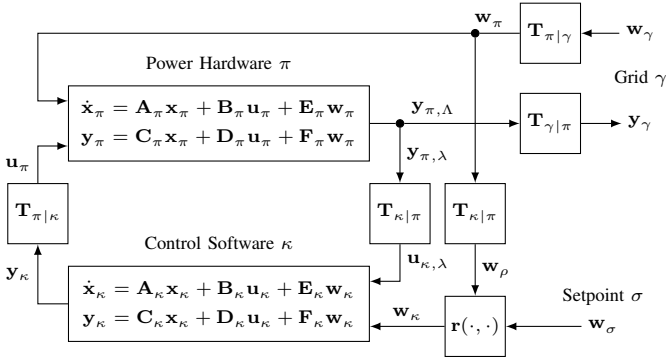

The proposed generic state-space model is obtained by combining the models of the power hardware, the control software, the transformations, and the reference calculation.
The corresponding block diagram is shown in \cref{fig:CIDER:model}.
Note that the proposed model is fully modular thanks to the transformation linking its blocks.
Therefore, each block can be modeled in a different frame of reference if desired.%
As previously mentioned, the grid is typically described in phase coordinates \cite{Jrn:PSE:SSA:1968:Laughton} or symmetrical components \cite{Jrn:PSE:SSA:1918:Fortescue}, and the control software in direct-quadrature-zero components \cite{Jrn:PSE:CA:1929:Park} or alpha-beta-gamma components \cite{Jrn:PSE:CA:1951:Duesterhoeft}.
The power hardware can be described in any of these reference frames.


\subsection{Harmonic-Domain State-Space Model}
\label{sec:CIDER:harmonic}

Recall that all matrix and vector quantities introduced in \cref{sec:CIDER:time} are time-periodic with period $T$.
Therefore, they can be written as Fourier series.
Namely
\begin{alignat}{2}
	    \mathbf{x}(t)
    &=	\sum\limits_{h\in\harmonics}\mathbf{X}_{h}\Exp{j h 2\pi f_{1} t}
    &,~ \text{etc.}\\
	    \mathbf{A}(t)
    &=  \sum\limits_{h\in\harmonics}\mathbf{A}_{h}\Exp{j h 2\pi f_{1} t}
    &,~ \text{etc.}	
\end{alignat}
where $f_{1}=\frac{1}{T}$ is the \emph{fundamental frequency} and $h\in\harmonics\subset\mathbb{Z}$ are the \emph{harmonic orders}.
As known from Fourier analysis, the multiplication of two waveforms in time domain corresponds to the convolution of their spectra in frequency domain:
\begin{equation}
                    \mathbf{A}(t)\mathbf{x}(t)
    \leftrightarrow \mathbf{A}(f)*\mathbf{X}(f)
    =               \hat{\mathbf{A}}\hat{\mathbf{X}}
\end{equation}
where $\hat{\mathbf{A}}$ is a Toeplitz matrix of the Fourier coefficients $\mathbf{A}_{h}$, and $\hat{\mathbf{X}}$ is column vector of the Fourier coefficients $\mathbf{X}_{h} $\cite{Ths:CSE:LTP:1991:Wereley}:
\begin{align}
	    \hat{\mathbf{A}}	&:~	\hat{\mathbf{A}}_{\mathit{mn}}=\mathbf{A}_{h},~m,n\in\mathbb{N},~h=m-n\in\harmonics\\
	\hat{\mathbf{X}}	&=	\col_{h\in\harmonics}(\mathbf{X}_{h})
\end{align}
Unless the associated signals are band-limited, such matrices and vectors are of infinite size.
In practice, only the harmonics up to a certain maximum order $h_{\mathit{max}}$ are considered%
\footnote{%
    Standards for voltage and power quality typically account for harmonics up to order 20-25 (i.e., 1.0-1.5 kHz) \cite{Std:BSI-EN-50160:2000}.
}.
Hence, the said Toeplitz matrices and column vectors are of finite size.

Consider the subsystem composed of the power hardware, control software, and the transforms which connect them (i.e., excluding the parts related to grid and reference calculation in \cref{fig:CIDER:model}).
The time-domain state-space models \eqref{eq:pwr:proc:time}--\eqref{eq:pwr:meas:time} and \eqref{eq:ctrl:proc:time}--\eqref{eq:ctrl:meas:time} can be formulated in the frequency domain
\begin{align}
        j\hat{\boldsymbol{\Omega}}_{\pwr}\hat{\mathbf{X}}_{\pwr}
    &=      \hat{\mathbf{A}}_{\pwr}\hat{\mathbf{X}}_{\pwr}
        +   \hat{\mathbf{B}}_{\pwr}\hat{\mathbf{U}}_{\pwr}
        +   \hat{\mathbf{E}}_{\pwr}\hat{\mathbf{W}}_{\pwr}
    \label{eq:pwr:proc:freq}\\
        \hat{\mathbf{Y}}_{\pwr}
    &=      \hat{\mathbf{C}}_{\pwr}\hat{\mathbf{X}}_{\pwr}
        +   \hat{\mathbf{D}}_{\pwr}\hat{\mathbf{U}}_{\pwr}
        +   \hat{\mathbf{F}}_{\pwr}\hat{\mathbf{W}}_{\pwr}
    \label{eq:pwr:meas:freq}\\
        j\hat{\boldsymbol{\Omega}}_{\ctrl}\hat{\mathbf{X}}_{\ctrl}
    &=      \hat{\mathbf{A}}_{\ctrl}\hat{\mathbf{X}}_{\ctrl}
        +   \hat{\mathbf{B}}_{\ctrl}\hat{\mathbf{U}}_{\ctrl}
        +   \hat{\mathbf{E}}_{\ctrl}\hat{\mathbf{W}}_{\ctrl}
    \label{eq:ctrl:proc:freq}\\
        \hat{\mathbf{Y}}_{\ctrl}
    &=      \hat{\mathbf{C}}_{\ctrl}\hat{\mathbf{X}}_{\ctrl}
        +   \hat{\mathbf{D}}_{\ctrl}\hat{\mathbf{U}}_{\ctrl}
        +   \hat{\mathbf{F}}_{\ctrl}\hat{\mathbf{W}}_{\ctrl}
    \label{eq:ctrl:meas:freq}
\end{align}
where the matrices $\hat{\boldsymbol{\Omega}}_{\pwr}$ and $\hat{\boldsymbol{\Omega}}_{\ctrl}$ are given by
\begin{align}
        \hat{\boldsymbol{\Omega}}_{\pwr}
    &=  2\pi f_{1} \diag_{h\in\harmonics}(h\cdot\mathbf{1}_{\pwr})
    \\
        \hat{\boldsymbol{\Omega}}_{\ctrl}
    &=  2\pi f_{1} \diag_{h\in\harmonics}(h\cdot\mathbf{1}_{\ctrl})
\end{align}
The time-domain transformations \eqref{eq:trafo:pwr-to-ctrl:time}--\eqref{eq:trafo:ctrl-to-pwr:time} can analogously be formulated in the frequency domain as
\begin{align}
        \hat{\mathbf{U}}_{\ctrl}
    &=  \hat{\mathbf{T}}^{+}_{\ctrl|\pwr}\hat{\mathbf{Y}}_{\pwr}
    \label{eq:trafo:pwr-to-ctrl:freq}\\
        \hat{\mathbf{U}}_{\pwr}
    &=  \hat{\mathbf{T}}_{\pwr|\ctrl}\hat{\mathbf{Y}}_{\ctrl}
    \label{eq:trafo:ctrl-to-pwr:freq}
\end{align}

Equations \eqref{eq:pwr:proc:freq}--\eqref{eq:pwr:meas:freq} and \eqref{eq:ctrl:proc:freq}--\eqref{eq:ctrl:meas:freq} can be combined into
\begin{align}
        j\hat{\boldsymbol{\Omega}}\hat{\mathbf{X}}
    &=      \hat{\mathbf{A}}\hat{\mathbf{X}}
        +   \hat{\mathbf{B}}\hat{\mathbf{U}}
        +   \hat{\mathbf{E}}\hat{\mathbf{W}}
    \label{eq:CIDER:proc:open}\\
        \hat{\mathbf{Y}}
    &=      \hat{\mathbf{C}}\hat{\mathbf{X}}
        +   \hat{\mathbf{D}}\hat{\mathbf{U}}
        +   \hat{\mathbf{F}}\hat{\mathbf{W}}
    \label{eq:CIDER:meas:open}
\end{align}
where
\begin{align}
	    \hat{\mathbf{X}}
    &=	\col(\hat{\mathbf{X}}_{\pwr},\hat{\mathbf{X}}_{\ctrl})
    \label{eq:CIDER:vars}\\
	    \hat{\mathbf{A}}
    &=  \diag(\hat{\mathbf{A}}_{\pwr},\hat{\mathbf{A}}_{\ctrl})
    \label{eq:CIDER:params}
\end{align}
The remaining vectors (i.e., $\hat{\mathbf{Y}}$ etc.) and matrices (i.e., $\hat{\mathbf{B}}$ etc.) are defined analogously.
Combining equations \eqref{eq:trafo:pwr-to-ctrl:freq}--\eqref{eq:trafo:ctrl-to-pwr:freq} yields
\begin{equation}
    \hat{\mathbf{U}} = \hat{\mathbf{T}}\mathbf{\hat{Y}}
    \label{eq:CIDER:trafo}
\end{equation}
where
\begin{equation}
        \hat{\mathbf{T}}
    =   \begin{bmatrix}
                \mathbf{0}
            &   \hat{\mathbf{T}}_{\pwr|\ctrl}\\
                \hat{\mathbf{T}}^{+}_{\ctrl|\pwr}
            &   \mathbf{0} 
        \end{bmatrix}
\end{equation}
One can interpret \eqref{eq:CIDER:proc:open}--\eqref{eq:CIDER:meas:open} as the open-loop model of the subsystem composed of power hardware and control software excl. reference calculation, and \eqref{eq:CIDER:trafo} as the associated feedback control law.
In order to obtain the closed-loop model, substitute \eqref{eq:CIDER:trafo} into \eqref{eq:CIDER:proc:open}--\eqref{eq:CIDER:meas:open} and solve for $\hat{\mathbf{X}}$ and $\hat{\mathbf{Y}}$:
\begin{align}
        j\Hat{\boldsymbol{\Omega}}\Hat{\mathbf{X}}
    &=      \Tilde{\mathbf{A}}\Hat{\mathbf{X}}
        +   \Tilde{\mathbf{E}}\Hat{\mathbf{W}}
    \label{eq:CIDER:proc:closed}\\
            \Hat{\mathbf{Y}}
    &=      \Tilde{\mathbf{C}}\Hat{\mathbf{X}}
        +   \Tilde{\mathbf{F}}\Hat{\mathbf{W}}
    \label{eq:CIDER:meas:closed}
\end{align}
where the matrices $\Tilde{\mathbf{A}}$, $\Tilde{\mathbf{C}}$, $\Tilde{\mathbf{E}}$, and $\Tilde{\mathbf{F}}$ are given by
\begin{align}
        \Tilde{\mathbf{A}}
    &=  \Hat{\mathbf{A}}+\Hat{\mathbf{B}}(\diag(\mathbf{1})-\Hat{\mathbf{T}}\Hat{\mathbf{D}})^{-1}\Hat{\mathbf{T}}\Hat{\mathbf{C}}
    \label{eq:CIDER:system}\\
        \Tilde{\mathbf{C}}
    &=  (\diag(\mathbf{1})-\Hat{\mathbf{D}}\Hat{\mathbf{T}})^{-1}\Hat{\mathbf{C}}
    \label{eq:CIDER:output}\\
        \Tilde{\mathbf{E}}
    &=  \Hat{\mathbf{E}}+\Hat{\mathbf{B}}(\diag(\mathbf{1})-\Hat{\mathbf{T}}\Hat{\mathbf{D}})^{-1}\Hat{\mathbf{T}}\Hat{\mathbf{F}}
    \label{eq:CIDER:input}\\
        \Tilde{\mathbf{F}}
    &=  (\diag(\mathbf{1})-\Hat{\mathbf{D}}\Hat{\mathbf{T}})^{-1}\Hat{\mathbf{F}}
    \label{eq:CIDER:feedthrough}
\end{align}
and $\diag(\mathbf{1})$ is an identity matrix of suitable size.
Naturally, the matrices \eqref{eq:CIDER:system}--\eqref{eq:CIDER:feedthrough} can only be computed if the inverses of $\diag(\mathbf{1})-\Hat{\mathbf{T}}\Hat{\mathbf{D}}$ and $\diag(\mathbf{1})-\Hat{\mathbf{D}}\Hat{\mathbf{T}}$ exist.
There is no general guarantee for this.
However, since the terms in question depend only on known parameters, one can assess offline (i.e., before the \HPF study) whether these conditions hold.
In this respect, the following hypothesis is made:
\begin{Hypothesis}\label{hyp:closed-loop:system}
    The closed-loop model \eqref{eq:CIDER:proc:closed}--\eqref{eq:CIDER:meas:closed} exists (i.e., $\diag(\mathbf{1})-\Hat{\mathbf{T}}\Hat{\mathbf{D}}$ and $\diag(\mathbf{1})-\Hat{\mathbf{D}}\Hat{\mathbf{T}}$ are invertible).
\end{Hypothesis}
\noindent
Solve \eqref{eq:CIDER:proc:closed} for $\Hat{\mathbf{X}}$ and substitute the result into \eqref{eq:CIDER:meas:closed} in order to obtain the closed-loop transfer function from $\Hat{\mathbf{W}}$ to $\Hat{\mathbf{Y}}$:
\begin{equation}
        \Hat{\mathbf{Y}}
    =   \Hat{\mathbf{G}}\Hat{\mathbf{W}},~
        \Hat{\mathbf{G}}
    =      \Tilde{\mathbf{C}}(j\Hat{\boldsymbol{\Omega}}-\Tilde{\mathbf{A}})^{-1}\Tilde{\mathbf{E}}
        +   \Tilde{\mathbf{F}}
    \label{eq:CIDER:transfer}
\end{equation}
$\Hat{\mathbf{G}}$ is the \emph{closed-loop gain}.
The existence of the closed-loop model alone (i.e., \cref{hyp:closed-loop:system}) does not guarantee the existence of the closed-loop gain.
In order for this to be the case, the term $j\Hat{\boldsymbol{\Omega}}-\Tilde{\mathbf{A}}$ needs to be invertible.
For the sake of rigour and transparency, this hypothesis is explicitly declared:
\begin{Hypothesis}\label{hyp:closed-loop:gain}
    The closed-loop transfer function \eqref{eq:CIDER:transfer} exists (i.e., the matrix $j\Hat{\boldsymbol{\Omega}}-\Tilde{\mathbf{A}}$ is invertible).
\end{Hypothesis}

Recall \eqref{eq:CIDER:vars}--\eqref{eq:CIDER:params} and write \eqref{eq:CIDER:transfer} in block form:
\begin{equation}
        \begin{bmatrix}
            \hat{\mathbf{Y}}_{\pwr}\\
            \hat{\mathbf{Y}}_{\ctrl}
        \end{bmatrix}
    =   \begin{bmatrix}
                \Hat{\mathbf{G}}_{\pwr\pwr}
            &   \Hat{\mathbf{G}}_{\pwr\ctrl}\\
                \Hat{\mathbf{G}}_{\ctrl\pwr}
            &   \Hat{\mathbf{G}}_{\ctrl\ctrl}
        \end{bmatrix}
        \begin{bmatrix}
            \hat{\mathbf{W}}_{\pwr}\\
            \hat{\mathbf{W}}_{\ctrl}
        \end{bmatrix}
    \label{eq:CIDER:transfer:block-form}
\end{equation}

In order to derive the closed-loop transfer function of the entire \CIDER (i.e., from $\hat{\mathbf{W}}_{\grid}$ to $\hat{\mathbf{Y}}_{\grid}$), the reference calculation needs to be included via the term $\hat{\mathbf{W}}_{\ctrl}$.
Recall that the time-domain function $\mathbf{r}(\cdot,\cdot)$ in \eqref{eq:ref:time} is in general nonlinear.
Hence, finding a corresponding relation in frequency domain may not be straightforward.
However, one can presume the following:
\begin{Hypothesis}\label{hyp:reference}
    There exists a differentiable function $\hat{\mathbf{R}}(\cdot,\cdot)$, which approximates $\mathbf{r}(\cdot,\cdot)$ in the harmonic domain:
    \begin{equation}
                \Hat{\mathbf{W}}_{\ctrl}
        \approx \Hat{\mathbf{R}}(\Hat{\mathbf{W}}_{\refr},\Hat{\mathbf{W}}_{\spt})
        \label{eq:CIDER:ref}
    \end{equation}
\end{Hypothesis}
\noindent
Differentiability of $\hat{\mathbf{R}}(\cdot,\cdot)$ is needed for the numerical solution of the \HPF equations%
\footnote{%
    The Newton-Raphson method requires the calculation of a Jacobian matrix in each iteration.
    Naturally, the Jacobian matrix exists only if the involved functions are differentiable.
}.
This will be discussed further shortly.
Finally, it is worth noting that $\hat{\mathbf{R}}(\cdot,\cdot)$ can be nonlinear.

In line with \eqref{eq:trafo:pwr-to-ref:time}, one finds that
\begin{equation}
        \Hat{\mathbf{W}}_{\refr}
    =   \Hat{\mathbf{T}}_{\ctrl|\pwr}\Hat{\mathbf{W}}_{\pwr}
    \label{eq:trafo:pwr-to-ref:freq}
\end{equation}
Through substitution of \eqref{eq:CIDER:ref} and \eqref{eq:trafo:pwr-to-ref:freq} into \eqref{eq:CIDER:transfer:block-form}, one obtains
\begin{equation}
        \Hat{\mathbf{Y}}_{\pwr}(\Hat{\mathbf{W}}_{\spt},\Hat{\mathbf{W}}_{\pwr})
    =       \Hat{\mathbf{G}}_{\pwr\pwr}\Hat{\mathbf{W}}_{\pwr}
        +   \Hat{\mathbf{G}}_{\pwr\ctrl}
            \Hat{\mathbf{R}}(\Hat{\mathbf{W}}_{\spt},\Hat{\mathbf{T}}_{\ctrl|\pwr}\Hat{\mathbf{W}}_{\pwr})
    \label{eq:CIDER:transfer:inner}
\end{equation}
In order to obtain the closed-loop transfer function w.r.t. the grid quantities $\hat{\mathbf{W}}_{\grid}$ and $\hat{\mathbf{Y}}_{\grid}$, the coordinate transformations between grid and power hardware need to be considered.%
From \eqref{eq:trafo:grd-to-pwr:time}--\eqref{eq:trafo:pwr-to-grd:time}, it follows that
\begin{align}
        \Hat{\mathbf{W}}_{\pwr}
    &=  \Hat{\mathbf{T}}_{\pwr|\grid}\Hat{\mathbf{W}}_{\grid}
    \label{eq:trafo:grd-to-pwr:freq}\\
        \Hat{\mathbf{Y}}_{\grid}
    &=  \Hat{\mathbf{T}}^{+}_{\grid|\pwr}\Hat{\mathbf{Y}}_{\pwr}
    \label{eq:trafo:pwr-to-grd:freq}
\end{align}
Combining \eqref{eq:CIDER:transfer:inner}--\eqref{eq:trafo:pwr-to-grd:freq} yields the desired transfer function:
\begin{equation}
        \Hat{\mathbf{Y}}_{\grid}(\Hat{\mathbf{W}}_{\grid},\Hat{\mathbf{W}}_{\spt})
    =   \Hat{\mathbf{T}}_{\grid|\pwr}^{+}
        \Hat{\mathbf{Y}}_{\pwr}
        (
            \Hat{\mathbf{T}}_{\pwr|\grid}\Hat{\mathbf{W}}_{\grid},
            \Hat{\mathbf{W}}_{\spt}
        )
    \label{eq:CIDER:transfer:outer}
\end{equation}
Recall from \eqref{eq:grid:disturbance:time}--\eqref{eq:setpoint:time} that this generic function can represent a grid-forming or a grid-following \CIDER (i.e., depending on which electrical quantities $\Hat{\Y}_{\grid}$, $\Hat{\mathbf{W}}_{\grid}$, and $\Hat{\mathbf{W}}_{\spt}$ correspond to).

As will be shown shortly, the partial derivative of $\Hat{\mathbf{Y}}_{\grid}$ w.r.t. $\Hat{\mathbf{W}}_{\grid}$ is needed for the numerical solution of the \HPF problem.
Note that $\Hat{\Y}_{\pwr}(\cdot,\cdot)$ in \eqref{eq:CIDER:transfer:outer} and $\Hat{\mathbf{R}}(\cdot,\cdot)$ in \eqref{eq:CIDER:transfer:inner} are differentiable (the former is a linear function, the latter due to \cref{hyp:reference}).
Hence, the chain rule can be applied, which yields
\begin{align}
        \partial_{\grid}\Hat{\mathbf{Y}}_{\grid}
        (\Hat{\mathbf{W}}_{\grid},\Hat{\mathbf{W}}_{\spt})
    &=  \Hat{\mathbf{T}}_{\grid|\pwr}^{+}
        \partial_{\pwr}\Hat{\mathbf{Y}}_{\pwr}
        (\Hat{\mathbf{T}}_{\pwr|\grid}\Hat{\mathbf{W}}_{\grid},\Hat{\mathbf{W}}_{\spt})
        \Hat{\mathbf{T}}_{\pwr|\grid}
    \\
        \partial_{\pwr}\Hat{\mathbf{Y}}_{\pwr}
        (\Hat{\mathbf{W}}_{\pwr},\Hat{\mathbf{W}}_{\spt})
    &=  \left[
        \begin{aligned}
            &\Hat{\mathbf{G}}_{\pwr\pwr}\\
            +&\Hat{\mathbf{G}}_{\pwr\ctrl}\partial_{\refr}\Hat{\mathbf{R}}
            (\Hat{\mathbf{T}}_{\ctrl|\pwr}\Hat{\mathbf{W}}_{\pwr},\Hat{\mathbf{W}}_{\spt})
            \Hat{\mathbf{T}}_{\ctrl|\pwr}
        \end{aligned}
        \right.
\end{align}
\noindent
where $\partial_{\grid}$, $\partial_{\pwr}$, and $\partial_{\refr}$ denote the partial derivatives w.r.t. $\Hat{\mathbf{W}}_{\grid}$, $\Hat{\mathbf{W}}_{\pwr}$, and $\Hat{\mathbf{W}}_{\refr}$, respectively.

\section{Algorithm for Harmonic Power-Flow Study}
\label{sec:HPF}


\subsection{Mathematical Formulation of the Problem}
\label{sec:HPF:problem}

The \HPF problem is obtained by formulating the mismatch equations between the models of the \CIDER[s] and the grid.
Without loss of generality, the nodes $\nodes$ are partitioned as
\begin{equation}
    \nodes=\formers\cup\followers,~\formers\cap\followers=\emptyset
\end{equation}
where $\formers$ and $\followers$ are the points of connection of grid-forming and grid-following \CIDER[s], respectively.
If there are any zero-injection nodes (i.e., without resources), they can be eliminated via Kron reduction \cite{Jrn:CT:NM:2019:Kettner}.

From the point of view of the grid, the nodal equations are given by the hybrid parameters \eqref{eq:nodes:hybrid:harmonics}:
\begin{align}
        \Hat{\mathbf{V}}_{\formers}(\Hat{\mathbf{I}}_{\formers},\Hat{\mathbf{V}}_{\followers})
    &=      \Hat{\mathbf{H}}_{\formers\times\formers}\Hat{\mathbf{I}}_{\formers}
        +   \Hat{\mathbf{H}}_{\formers\times\followers}\Hat{\mathbf{V}}_{\followers}
    \label{eq:grid:form}
    \\
        \Hat{\mathbf{I}}_{\followers}(\Hat{\mathbf{I}}_{\formers},\Hat{\mathbf{V}}_{\followers})
    &=      \Hat{\mathbf{H}}_{\followers\times\formers}\Hat{\mathbf{I}}_{\formers}
        +   \Hat{\mathbf{H}}_{\followers\times\followers}\Hat{\mathbf{V}}_{\followers}
    \label{eq:grid:follow}
\end{align}
From the point of view of the \CIDER[s], these nodal equations are established via the closed-loop transfer function \eqref{eq:CIDER:transfer:outer}.
Recalling the definitions of the setpoint disturbance $\mathbf{w}_{\spt}$ \eqref{eq:setpoint:time}, grid disturbance $\mathbf{w}_{\grid}$ \eqref{eq:grid:disturbance:time}, and grid output $\mathbf{y}_{\grid}$ \eqref{eq:grid:output:time}, one finds
\begin{alignat}{2}
        s\in\formers
    &:  ~
    &   \Hat{\mathbf{V}}_{s}(\Hat{\mathbf{I}}_{s},V_{\spt,s},f_{\spt,s})
    &=  \Hat{\mathbf{Y}}_{\grid,s}(\Hat{\mathbf{T}}_{\pwr|\grid}\Hat{\mathbf{I}}_{s},V_{\spt,s},f_{\spt,s})
    \label{eq:resource:form}\\
        r\in\followers
    &:  ~
    &   \Hat{\mathbf{I}}_{r}(\Hat{\mathbf{V}}_{r},S_{\spt,r})
    &=  \Hat{\mathbf{Y}}_{\grid,r}(\Hat{\mathbf{T}}_{\pwr|\grid}\Hat{\mathbf{V}}_{r},S_{\spt,r})
    \label{eq:resource:follow}
\end{alignat}
Note that \eqref{eq:resource:form} and \eqref{eq:resource:follow} are in accordance with \cref{def:CIDER:forming,def:CIDER:following} (i.e., grid-forming and grid-following behaviour).
Moreover, observe that transfer functions of the form \eqref{eq:resource:form} or \eqref{eq:resource:follow} can also be used to represent sources of harmonics other than \CIDER[s], such as conventional generators or loads.
The sole difference is that the transfer functions have to be derived from another suitable model or through system identification.
Please see Appendix~\ref{app:harmonics} for further details on this matter.

The mismatches between \eqref{eq:grid:form}--\eqref{eq:grid:follow} and \eqref{eq:resource:form}--\eqref{eq:resource:follow} must be zero in equilibrium (by definition):
\begin{align}
        \Delta\Hat{\mathbf{V}}_{\formers}
        (\Hat{\mathbf{I}}_{\formers},\Hat{\mathbf{V}}_{\followers},\mathbf{V}_{\spt},\mathbf{f}_{\spt})
    &=  \mathbf{0}
    \label{eq:residual:form}\\
        \Delta\Hat{\mathbf{I}}_{\followers}
        (\Hat{\mathbf{I}}_{\formers},\Hat{\mathbf{V}}_{\followers},\mathbf{S}_{\spt})
    &=  \mathbf{0}
    \label{eq:residual:follow}
\end{align}
where $\mathbf{V}_{\spt}$, $\mathbf{f}_\spt$, and $\mathbf{S}_{\spt}$ are column vectors built of $V_{\spt,s}$, $f_{\spt,s}$ ($s\in\formers$) and $S_{\spt,r}$ ($r\in\followers$), respectively.
In contrast to existing formulations (e.g., \cite{Jrn:PSE:SSA:1998:Smith,Jrn:PSE:SSA:2003:Herraiz,Jrn:PSE:SSA:2013:IEEE}), the so-called \emph{mismatch equations} \eqref{eq:residual:form}--\eqref{eq:residual:follow} are in hybrid rather than admittance form.
This reflects the grid-forming and grid-following behaviour.


\subsection{Numerical Solution via the Newton-Raphson Method}
\label{sec:HPF:solution}

\begin{algorithm}[t]
	\centering

{
\renewcommand{\arraystretch}{1.1}

\begin{algorithmic}
    \Procedure{\texttt{\HPF}}
    {%
        $\Delta\Hat{\mathbf{V}}_{\formers}(\cdot,\cdot,\cdot)$,
        $\Delta\Hat{\mathbf{I}}_{\followers}(\cdot,\cdot,\cdot)$,
        $\mathbf{V}_{\spt}$,
        $\mathbf{f}_{\spt}$,
        $\mathbf{S}_{\spt}$
    }
        \State
        {%
            \texttt{\# Initialization}
        }
        \State
        {%
            $\Hat{\I}_{\formers} \gets \mathbf{0}$
        }
        \State
        {%
            $\Hat{\V}_{\followers} \gets \texttt{flat\_start()}$
        }
        \While
        {$
            \max(|\Delta\Hat{\mathbf{V}}_{\formers}|,|\Delta\Hat{\mathbf{I}}_{\followers}|)
            \geqslant\epsilon
        $}
            \State{\texttt{\# Residuals}}
            \State
            {$
                        \Delta\Hat{\mathbf{V}}_{\formers}
                \gets   \Delta\Hat{\mathbf{V}}_{\formers}
                        (\Hat{\mathbf{I}}_{\formers},\Hat{\mathbf{V}}_{\followers},\mathbf{V}_{\spt},\mathbf{f}_{\spt})
            $}
            \State
            {$
                        \Delta\Hat{\mathbf{I}}_{\followers}
                \gets   \Delta\Hat{\mathbf{I}}_{\followers}
                        (\Hat{\mathbf{I}}_{\formers},\Hat{\mathbf{V}}_{\followers},\mathbf{S}_{\spt})
            $}
            \State{\texttt{\# Jacobian matrix}}
            \State
            {$
                        \Hat{\mathbf{J}}_{\formers\times\formers}
                \gets   \partial_{\formers}\Delta\Hat{\mathbf{V}}_{\formers}
                        (\Hat{\mathbf{I}}_{\formers},\Hat{\mathbf{V}}_{\followers},\mathbf{V}_{\spt},\mathbf{f}_{\spt})
            $}
            \State
            {$
                        \Hat{\mathbf{J}}_{\formers\times\followers}
                \gets   \partial_{\followers}\Delta\Hat{\mathbf{V}}_{\formers}
                        (\Hat{\mathbf{I}}_{\formers},\Hat{\mathbf{V}}_{\followers},\mathbf{V}_{\spt},\mathbf{f}_{\spt})
            $}
            \State
            {$
                        \Hat{\mathbf{J}}_{\followers\times\formers}
                \gets   \partial_{\formers}\Delta\Hat{\mathbf{I}}_{\followers}
                        (\Hat{\mathbf{I}}_{\formers},\Hat{\mathbf{V}}_{\followers},\mathbf{S}_{\spt})
            $}
            \State
            {$
                        \Hat{\mathbf{J}}_{\followers\times\followers}
                \gets   \partial_{\followers}\Delta\Hat{\mathbf{I}}_{\followers}
                        (\Hat{\mathbf{I}}_{\formers},\Hat{\mathbf{V}}_{\followers},\mathbf{S}_{\spt})
            $}
            \State{\texttt{\# Newton-Raphson iteration}}
            \State
            {$
                \begin{bmatrix}
                    \Delta\Hat{\mathbf{I}}_{\formers}\\
                    \Delta\Hat{\mathbf{V}}_{\followers}
                \end{bmatrix}
                \gets
                \begin{bmatrix}
                        \Hat{\mathbf{J}}_{\formers\times\formers}
                    &   \Hat{\mathbf{J}}_{\formers\times\followers}\\
                        \Hat{\mathbf{J}}_{\followers\times\formers}
                    &   \Hat{\mathbf{J}}_{\followers\times\followers}
                \end{bmatrix}^{-1}
                \begin{bmatrix}
                    \Delta\Hat{\mathbf{V}}_{\formers}\\
                    \Delta\Hat{\mathbf{I}}_{\followers}
                \end{bmatrix}
            $}
            \State
            {$
                \begin{bmatrix}
                    \Hat{\mathbf{I}}_{\formers}\\
                    \Hat{\mathbf{V}}_{\followers}
                \end{bmatrix}
                \gets
                    \begin{bmatrix}
                        \Hat{\mathbf{I}}_{\formers}\\
                        \Hat{\mathbf{V}}_{\followers}
                    \end{bmatrix}
                -   \begin{bmatrix}
                        \Delta\Hat{\mathbf{I}}_{\formers}\\
                        \Delta\Hat{\mathbf{V}}_{\followers}
                    \end{bmatrix}
            $}
        \EndWhile
    \EndProcedure
\end{algorithmic}
}
	\caption{Newton-Raphson solution of the \HPF problem.}
	\label{alg:newton-raphson}
\end{algorithm}

The \HPF problem is solved numerically via the Newton-Raphson method as described in \cref{alg:newton-raphson}.

In general, the \HPF problem may have multiple equilibrium points -- like any nonlinear problem.
If these equilibrium points lie in close proximity in the solution space, it is not evident to which solution the \HPF method will converge.
Notably, one or several of such neighbouring solutions may not be physically meaningful (i.e., analogous to solutions of the power-flow equations which lie on the lower portion of the well-known nose curve \cite{Bk:PSE:SSA:1998:VanCutsem}).
Hence, the choice of the initial point may affect which solution the numerical solver converges to, and whether this solution is physically meaningful.
Without any prior information (e.g., the solution of another \HPF study for a similar operating point), the initial point can be chosen as follows.
The injected currents of the nodes with grid-forming \CIDER[s] are initialized with 0.
The phase-to-ground voltages of the nodes with grid-following \CIDER[s] are initialized with a ``flat profile'', namely: the fundamental voltage is set to a pure positive sequence with magnitude 1~p.u. and phase equal to 0~rad, and the harmonic voltages are set to 0.
As will be discussed in Part~II, the proposed method is robust w.r.t. the choice of the initial point (i.e., it converges reliably even if the initial point lies far from the final solution).
The described procedure is preferred solely for its simplicity.

The Jacobian matrix has to be recomputed in each iteration of the Newton-Raphson method (as usual).
In the proposed formulation of the \HPF problem, most terms in the Jacobian matrix are constant, which reduces the computational intensity.
The partial derivatives of the grid model \eqref{eq:grid:form}--\eqref{eq:grid:follow}, which is linear, are the hybrid parameters:
\begin{align}
        \partial_{\formers}\Hat{\mathbf{V}}_{\formers}
        (\Hat{\I}_{\formers},\Hat{\V}_{\followers},\V_{\spt},\mathbf{f}_{\spt})
    &=  \Hat{\mathbf{H}}_{\formers\times\formers}\\
        \partial_{\followers}\Hat{\mathbf{V}}_{\formers}
        (\Hat{\I}_{\formers},\Hat{\V}_{\followers},\V_{\spt},\mathbf{f}_{\spt})
    &=  \Hat{\mathbf{H}}_{\formers\times\followers}\\
        \partial_{\formers}\Hat{\mathbf{I}}_{\followers}
        (\Hat{\I}_{\formers},\Hat{\V}_{\followers},\mathbf{S}_{\spt})
    &=  \Hat{\mathbf{H}}_{\followers\times\formers}\\
        \partial_{\followers}\Hat{\mathbf{I}}_{\followers}
        (\Hat{\I}_{\formers},\Hat{\V}_{\followers},\mathbf{S}_{\spt})
    &=  \Hat{\mathbf{H}}_{\followers\times\followers}
\end{align}
which only need to be updated if the electrical parameters or the topology of the grid change.
The partial derivatives of the \CIDER models \eqref{eq:resource:form}--\eqref{eq:resource:follow} are given by
\begin{align}
        \partial_{s}\Hat{\mathbf{V}}_{s}(\Hat{\mathbf{I}}_{s},V_{\spt,s},f_{\spt,s})
    &=  \partial_{\grid}\Hat{\mathbf{Y}}_{\grid,s}
        (
            \Hat{\mathbf{T}}_{\pwr|\grid}\Hat{\mathbf{I}}_{,s},
            V_{\spt,s},
            f_{\spt,s}
        )\\
        \partial_{r}\Hat{\mathbf{I}}_{r}(\Hat{\mathbf{V}}_{r},S_{\spt,r})
    &=  \partial_{\grid}\Hat{\mathbf{Y}}_{\grid,r}
        (
            \Hat{\mathbf{T}}_{\pwr|\grid}\Hat{\mathbf{I}}_{r},
            S_{\spt,r}
        )
\end{align}
Recall from \cref{sec:CIDER} that the reference calculation is the only block in the \CIDER model which may be nonlinear.
Hence, only the partial derivatives associated with resources for which the function $\Hat{\mathbf{R}}(\cdot,\cdot)$ in \eqref{eq:CIDER:ref} is actually nonlinear have to be updated in each iteration.
The partial derivatives of the other resources need to be calculated only once.

It is important to note that, in contrast to many existing \IHA methods (see \cref{sec:review}), \cref{alg:newton-raphson} is \emph{single-iterative} rather than \emph{double-iterative}.
This is thanks to the use of closed-loop transfer functions, which allow to incorporate the \CIDER behaviour \eqref{eq:resource:form}--\eqref{eq:resource:follow} directly into the nodal mismatch equations.
As a result, a single-iterative calculation at the system level suffices to solve the \HPF problem%
\footnote{%
    The standard approach is a double-iterative calculation with nested loops: one loop each at system and resource level, respectively.
}.

\section{Conclusions}
\label{sec:conclusion}

In this paper, a method for the \HPF study of three-phase power grids with \CIDER[s] has been proposed.
The underlying modelling framework is based on polyphase circuit theory and \LTP systems theory.
More precisely, a generic three-phase grid is described by hybrid nodal equations, and the \CIDER[s] by closed-loop transfer functions.
In this way, the creation and propagation of harmonics through the individual resources and the entire grid can be modeled with high fidelity.
The transfer functions are derived from a \CIDER model which is generic w.r.t. the control law (i.e., grid-forming or grid-following mode) and modular w.r.t. the components (i.e., power hardware, control software, and reference calculation).
The \HPF problem is defined by the mismatches between the models of the grid and the resources, which are zero in equilibrium.
This system of nonlinear equations can be solved efficiently via the Newton-Raphson method: a single-iterative algorithm is sufficient, and many terms in the Jacobian matrix are constant.
In future works, the solvability of the \HPF problem will be further investigated, in order to assess the harmonic stability property of the system.

\appendices

\section{Modelling of Sources of Harmonics\\ Other than Converter-Interfaced Resources}
\label{app:harmonics}

Naturally, harmonics may originate from sources which are not \CIDER[s], such as conventional resources or upstream and downstream power grids (i.e., background harmonics).
Indeed, the proposed approach can accommodate such sources of harmonics which are not converter-interfaced.
Namely -- much like \CIDER[s] -- they can be represented by transfer functions in the harmonic domain.
In this respect, the sole prerequisite is that such transfer functions can somehow be obtained -- i.e., either from a suitable model or via system identification.

For instance, a harmonic \emph{Th{\'e}venin Equivalent} (\TE) or \emph{Norton Equivalent} (\NE) can be used.
Let $m\in\nodes$ be a node at which a non-\CIDER source of harmonics is located.
If a \TE is used, the injected current is given by
\begin{equation}
    \Hat{\I}_{m} = \Hat{\Z}_{\TE,m}^{-1}(\Hat{\V}_{m}-\hat{\V}_{\TE,m})
\end{equation}
where $\Hat{\V}_{\TE}$ and $\Hat{\Z}_{\TE}$ are the harmonic voltage source and harmonic impedance, respectively, of the \TE.
If a \NE is used instead, the injected current is given by
\begin{equation}
    \Hat{\I}_{m} = \Hat{\I}_{\NE,m} - \Hat{\Y}_{\NE,m}\Hat{\V}_{m}
\end{equation}
where $\Hat{\I}_{\NE}$ and $\Hat{\Y}_{\TE}$ are the harmonic current source and harmonic admittance, respectively, of the \NE.
In these cases, the generation of harmonics is represented by the equivalent voltage or current sources, and the coupling between harmonics by the equivalent impedances or admittances (i.e., by the off-diagonal blocks of these matrices).



\bibliographystyle{IEEEtran}
\bibliography{Bibliography}

\begin{thebibliography}{10}
\providecommand{\url}[1]{#1}
\csname url@samestyle\endcsname
\providecommand{\newblock}{\relax}
\providecommand{\bibinfo}[2]{#2}
\providecommand{\BIBentrySTDinterwordspacing}{\spaceskip=0pt\relax}
\providecommand{\BIBentryALTinterwordstretchfactor}{4}
\providecommand{\BIBentryALTinterwordspacing}{\spaceskip=\fontdimen2\font plus
\BIBentryALTinterwordstretchfactor\fontdimen3\font minus
  \fontdimen4\font\relax}
\providecommand{\BIBforeignlanguage}[2]{{%
\expandafter\ifx\csname l@#1\endcsname\relax
\typeout{** WARNING: IEEEtran.bst: No hyphenation pattern has been}%
\typeout{** loaded for the language `#1'. Using the pattern for}%
\typeout{** the default language instead.}%
\else
\language=\csname l@#1\endcsname
\fi
#2}}
\providecommand{\BIBdecl}{\relax}
\BIBdecl

\bibitem{Jrn:PSE:PEC:2004:Blaabjerg}
F.~Blaabjerg, Z.~Chen, and S.~B. Kjaer, ``Power electronics as efficient
  interface in dispersed power generation systems,'' \emph{IEEE Trans. Power
  Electron.}, vol.~19, no.~5, pp. 1184--1194, 2004.

\bibitem{Bk:PSE:2016:Milano}
F.~Milano, Ed., \emph{Advances in Power System Modelling, Control, and
  Stability Analysis}.\hskip 1em plus 0.5em minus 0.4em\relax Stevenage, ENG,
  UK: IET, 2016.

\bibitem{Rep:PSE:C:2011:CIGRE}
C.~D'Adamo \emph{et~al.}, ``Development and operation of active distribution
  networks,'' CIGR{\'E}, Tech. Rep. 457, 2011.

\bibitem{Rep:PSE:SA:2018:Canizares}
C.~A. Ca{\~n}izares \emph{et~al.}, ``Microgrid stability definitions, analysis,
  and modeling,'' IEEE PES, Tech. Rep. PES-TR66, 2018.

\bibitem{Rep:PSE:SA:2020:Hatziargyriou}
N.~Hatziargyriou \emph{et~al.}, ``Stability definitions and characterization of
  dynamic behavior in systems with high penetration of power electronic
  interfaced technologies,'' IEEE PES, Tech. Rep. PES-TR77, 2020.

\bibitem{Jrn:PSE:PEC:2004:Enslin}
J.~Enslin and P.~Heskes, ``Harmonic interaction between a large number of
  distributed power inverters and the distribution network,'' \emph{IEEE Trans.
  Power Electron.}, vol.~19, no.~6, pp. 1586--1593, 2004.

\bibitem{Jrn:PSE:SSA:2017:Safargholi:Part1}
F.~Safargholi, K.~Malekian, and W.~Schufft, ``On the dominant harmonic source
  identification — part i: Review of methods,'' \emph{IEEE Trans. Power
  Del.}, vol.~33, no.~3, pp. 1268--1277, 2017.

\bibitem{Jrn:PSE:SSA:2017:Safargholi:Part2}
------, ``On the dominant harmonic source identification — part ii:
  Application and interpretation of methods,'' \emph{IEEE Trans. Power Del.},
  vol.~33, no.~3, pp. 1278--1287, 2017.

\bibitem{Bk:PSE:SSA:1997:Arrillaga}
J.~Arrillaga \emph{et~al.}, \emph{Power System Harmonic Analysis}.\hskip 1em
  plus 0.5em minus 0.4em\relax Hoboken, NJ, USA: Wiley, 1997.

\bibitem{Rep:PSE:SA:2004:Kundur}
P.~S. Kundur \emph{et~al.}, ``Definition and classification of power system
  stability,'' \emph{IEEE Trans. Power Syst.}, vol.~19, no.~3, pp. 1387--1401,
  2004.

\bibitem{Jrn:PSE:SSA:1987:Arrillaga}
J.~Arrillaga \emph{et~al.}, ``Comparison of steady-state and dynamic models for
  the calculation of {AC}/{DC} system harmonics,'' \emph{IEE Proc. C -- Gener.
  Transm. Distrib.}, vol. 134, no.~1, pp. 31--37, 1987.

\bibitem{Jrn:PSE:SSA:1998:Smith}
B.~C. Smith \emph{et~al.}, ``A review of iterative harmonic analysis for
  {AC}/{DC} power systems,'' \emph{IEEE Trans. Power Del.}, vol.~13, no.~1, pp.
  180--185, 1998.

\bibitem{Jrn:PSE:SSA:2003:Herraiz}
S.~Herraiz \emph{et~al.}, ``Review of harmonic load-flow formulations,''
  \emph{IEEE Trans. Power Del.}, vol.~18, no.~3, pp. 1079--1087, 2003.

\bibitem{Jrn:PSE:SSA:2013:IEEE}
A.~Medina \emph{et~al.}, ``Harmonic analysis in frequency and time domain,''
  \emph{IEEE Trans. Power Del.}, vol.~28, no.~3, pp. 1813--1821, 2013.

\bibitem{Jrn:PSE:SSA:2019:Wang}
X.~Wang and F.~Blaabjerg, ``Harmonic stability in power-electronic-based power
  systems: Concept, modeling, and analysis,'' \emph{IEEE Trans. Smart Grid},
  vol.~10, no.~3, pp. 2858--2870, 2019.

\bibitem{Bk:CT:NA:1975:Chua}
L.~O. Chua and P.-M. Lin, \emph{Computer-Aided Analysis of Electronic
  Circuits}.\hskip 1em plus 0.5em minus 0.4em\relax Upper Saddle River, NJ,
  USA: Prentice Hall, 1975.

\bibitem{Bk:CT:NA:1975:Dimo}
P.~Dimo, \emph{Nodal Analysis of Power Systems}.\hskip 1em plus 0.5em minus
  0.4em\relax Tunbridge Wells, UK: Abacus Press, 1975.

\bibitem{Jrn:PSA:TA:1969:Dommel}
H.~W. Dommel, ``Digital computer solution of electromagnetic transients in
  single- and multiphase networks,'' \emph{IEEE Trans. Power App. Syst.},
  no.~4, pp. 388--399, 1969.

\bibitem{Rep:CT:NA:1973:Nagel}
L.~W. Nagel and D.~O. Pederson, ``{SPICE}: Simulation program with integrated
  circuit emphasis,'' UC Berkeley, Tech. Rep. UCB M382, 1973.

\bibitem{Jrn:CT:NA:1975:Ho}
C.-W. Ho, A.~E. R{\"u}hli, and P.~A. Brennan, ``The modified nodal approach to
  network analysis,'' \emph{IEEE Trans. Circuits Syst.}, vol.~22, no.~6, pp.
  504--509, 1975.

\bibitem{Jrn:PSA:TA:2007:Mahseredjian}
J.~Mahseredjian \emph{et~al.}, ``On a new approach for the simulation of
  transients in power systems,'' \emph{Elect. Power Syst. Research}, vol.~77,
  no.~11, pp. 1514--1520, 2007.

\bibitem{Jrn:PSA:TA:2017:Gu}
Y.~Gu, N.~Bottrell, and T.~C. Green, ``Reduced-order models for representing
  converters in power system studies,'' \emph{IEEE Trans. Power Electron.},
  vol.~33, no.~4, pp. 3644--3654, 2017.

\bibitem{Jrn:PSA:TA:2019:Todeschini}
G.~Todeschini, S.~Balasubramaniam, and P.~Igic, ``Time-domain modeling of a
  distribution system to predict harmonic interaction between {PV}
  converters,'' \emph{IEEE Trans. Sust. Energy}, vol.~10, no.~3, pp.
  1450--1458, 2019.

\bibitem{Jrn:PSE:SSA:1995:Arrillaga}
J.~Arrillaga \emph{et~al.}, ``The harmonic domain. a frame of reference for
  power system harmonic analysis,'' \emph{IEEE Trans. Power Syst.}, vol.~10,
  no.~1, pp. 433--440, 1995.

\bibitem{Jrn:PSE:SSA:2004:Arrillaga}
J.~Arrillaga, N.~R. Watson, and G.~N. Bathurst, ``A multifrequency power flow
  of general applicability,'' \emph{IEEE Trans. Power Del.}, vol.~19, no.~1,
  pp. 342--349, 2004.

\bibitem{Jrn:PSE:SSA:1982:Xia}
D.~Xia and G.~T. Heydt, ``Harmonic power flow studies. {Part I+II},''
  \emph{IEEE Trans. Power App. Syst.}, vol. 101, no.~6, pp. 1257--1270, Jun.
  1982.

\bibitem{Jrn:PSE:SSA:1991:Xu:1}
W.~Xu, J.~R. Mart{\'i}, and H.~W. Dommel, ``A multiphase harmonic load-flow
  solution technique,'' \emph{IEEE Trans. Power Syst.}, vol.~6, no.~1, pp.
  174--182, 1991.

\bibitem{Jrn:PSE:SSA:1993:Valcarel}
M.~Valc{\'a}rcel and J.~Garcia~Mayordomo, ``Harmonic power flow for unbalanced
  systems,'' \emph{IEEE Trans. Power Del.}, vol.~8, no.~4, pp. 2052--2059, Oct.
  1993.

\bibitem{Jrn:PSE:SSA:1968:Laughton}
M.~A. Laughton, ``Analysis of unbalanced polyphase networks by the method of
  phase coordinates. {Part 1}: System representation in phase frame of
  reference,'' \emph{Proc. IEE}, vol. 115, no.~8, pp. 1163--1172, 1968.

\bibitem{Jrn:PSE:SSA:1918:Fortescue}
C.~L. Fortescue, ``Method of symmetrical co-ordinates applied to the solution
  of polyphase networks,'' \emph{Trans. AIEE}, vol.~37, no.~2, pp. 1027--1140,
  1918.

\bibitem{Jrn:PSE:SSA:1991:Arrillaga}
J.~Arrillaga and C.~Callaghan, ``Three-phase {AC}/{DC} load and harmonic
  flows,'' \emph{IEEE Trans. Power Del.}, vol.~6, no.~1, pp. 238--244, 1991.

\bibitem{Jrn:PSE:SSA:2020:Chen}
D.~Chen and L.~Xiao, ``A novel fundamental and harmonics detection method based
  on state-space model for power electronics system,'' \emph{IEEE Access},
  vol.~8, pp. 170\,002--170\,012, 2020.

\bibitem{Ths:CSE:LTP:1991:Wereley}
N.~M. Wereley, ``Analysis and control of linear periodically time-varying
  systems,'' Ph.D. dissertation, MIT, Cambridge, MA, USA, 1991.

\bibitem{J:PSA:SSA:2003:Rico}
J.~Rico~Melgoza, M.~Madrigal~Mart{\'i}nez, and E.~Acha~Daza, ``Dynamic harmonic
  evolution using the extended harmonic domain,'' \emph{IEEE Trans. Power
  Del.}, vol.~18, no.~2, pp. 587--594, 2003.

\bibitem{Jrn:PSE:SSA:2017:Kwon}
J.~B. Kwon \emph{et~al.}, ``Harmonic interaction analysis in a grid-connected
  converter using harmonic state-space modeling,'' \emph{IEEE Trans. Power
  Electron.}, vol.~32, no.~9, pp. 6823--6835, 2016.

\bibitem{Jrn:PSE:SSA:2017:Sun}
J.~Sun and H.~Liu, ``Sequence impedance modeling of modular multilevel
  converters,'' \emph{IEEE Journal of Emerging and Selected Topics in Power
  Electronics}, vol.~5, no.~4, pp. 1427--1443, 2017.

\bibitem{Jrn:PSE:SSA:2018:Abbood}
H.~D. Abbood and A.~Benigni, ``Data-driven modeling of a commercial
  photovoltaic microinverter,'' \emph{Modelling Simulation Eng.}, 2018.

\bibitem{Cnf:PSE:SSA:2019:Liu}
Y.~Liu, Y.~Li, J.~Ren, S.~Wang, and L.~Li, ``A data-driven harmonic modeling
  method for electric vehicle charging stations,'' \emph{Proc. Int. Conf.
  Electricity Distrib. (CIRED), Madrid, Spain}, 2019.

\bibitem{Jrn:PSE:SSA:2020:Nduka}
O.~S. Nduka and A.~R. Ahmadi, ``Data-driven robust extended computer-aided
  harmonic power flow analysis,'' \emph{IET Gener. Transm. Distrib.}, vol.~14,
  no.~20, pp. 4398--4409, 2020.

\bibitem{Jrn:PSE:HA:1995:Semlyen}
A.~Semlyen and A.~Medina, ``Computation of the periodic steady-state in systems
  with nonlinear components using a hybrid time- and frequency-domain
  methodology,'' \emph{IEEE Trans. Power Syst.}, vol.~10, no.~3, pp.
  1498--1504, 1995.

\bibitem{Jrn:PSE:HA:2007:Wiechowski}
W.~Wiechowski \emph{et~al.}, ``Hybrid time/frequency-domain modelling of
  nonlinear components,'' in \emph{Proc. Int. Conf. Elect. Power Qual. Util.,
  Barcelona, CAT, ES}, 2007, pp. 1--6.

\bibitem{Jrn:TSG:TA:2019:Peng}
Y.~Peng, Z.~Shuai, X.~Liu, Z.~Li, J.~M. Guerrero, and Z.~J. Shen, ``Modeling
  and stability analysis of inverter-based microgrid under harmonic
  conditions,'' \emph{IEEE Trans. Smart Grid}, vol.~11, no.~2, pp. 1330--1342,
  2019.

\bibitem{Jrn:CT:NM:2019:Kettner}
A.~M. Kettner and M.~Paolone, ``On the properties of the compound nodal
  admittance matrix of polyphase power systems,'' \emph{IEEE Trans. Power
  Syst.}, vol.~34, no.~1, pp. 444--453, 2019.

\bibitem{Bk:PSE:SSA:1990:Arrillaga}
J.~Arrillaga and C.~P. Arnold, \emph{Computer Analysis of Power Systems}.\hskip
  1em plus 0.5em minus 0.4em\relax Chichester, SXW, UK: Wiley, 1990.

\bibitem{Bk:CT:NM:2016:Borghetti}
A.~Borghetti \emph{et~al.}, ``Telegrapher's equations for
  field-to-transmission-line interaction,'' in \emph{Advances in Power System
  Modelling, Control, and Stability Analysis}, F.~Milano, Ed.\hskip 1em plus
  0.5em minus 0.4em\relax Stevenage, ENG, UK: IET, 2016.

\bibitem{Jrn:CT:NM:2001:Gustavsen}
B.~Gustavsen and A.~Semlyen, ``Enforcing passivity for admittance matrices
  approximated by rational functions,'' \emph{IEEE Trans. Power Syst.},
  vol.~16, no.~1, pp. 97--104, Feb. 2001.

\bibitem{Jrn:CT:NM:1982:Alvarado}
F.~L. Alvarado, ``Formation of ${Y}$-node using the primitive ${Y}$-node
  concept,'' \emph{IEEE Trans. Power App. Syst.}, no.~12, pp. 4563--4571, 1982.

\bibitem{Jrn:PSE:C:2016:Xin}
Z.~Xin \emph{et~al.}, ``Grid-current-feedback control for ${LCL}$-filtered grid
  converters with enhanced stability,'' \emph{IEEE Trans. Power Electron.},
  vol.~32, no.~4, pp. 3216--3228, 2016.

\bibitem{Jrn:PSE:C:2012:Guerrero:1}
J.~M. Guerrero \emph{et~al.}, ``Advanced control architectures for intelligent
  microgrids, part i: Decentralized and hierarchical control,'' \emph{IEEE
  Trans. Ind. Electron.}, vol.~60, no.~4, pp. 1254--1262, 2012.

\bibitem{Jrn:PSE:C:2012:Guerrero:2}
J.~Guerrero \emph{et~al.}, ``Advanced control architectures for intelligent
  microgrids, part ii: Power quality, energy storage, and ac/dc microgrids,''
  \emph{IEEE Trans. Ind. Electron.}, vol.~60, no.~4, pp. 1263--1270, 2012.

\bibitem{Cnf:PSE:SSA:2008:Love}
G.~N. Love and A.~R. Wood, ``Harmonic state-space model of power electronics,''
  in \emph{Proc. Int. Conf. Harmon. Qual. Power, Wollongong, NSW, AU}, 2008,
  pp. 1--6.

\bibitem{Jrn:PSE:CA:1951:Duesterhoeft}
W.~C. Duesterhoeft, M.~W. Schulz, and E.~Clarke, ``Determination of
  instantaneous currents and voltages by means of $\alpha$, $\beta$, and $0$
  components,'' \emph{Trans. AIEE}, vol.~70, no.~2, pp. 1248--1255, 1951.

\bibitem{Jrn:PSE:CA:1929:Park}
R.~H. Park, ``Two-reaction theory of synchronous machines,'' \emph{Trans.
  AIEE}, vol.~48, no.~3, pp. 716--727, 1929.

\bibitem{Std:BSI-EN-50160:2000}
``Voltage characteristics of electricity supplied by public distribution
  networks,'' British Standards Institution, London, UK, Std. BS-EN-50160:2000,
  2000.

\bibitem{Bk:PSE:SSA:1998:VanCutsem}
T.~Van~Cutsem and C.~Vournas, \emph{Voltage Stability of Electric Power
  Systems}.\hskip 1em plus 0.5em minus 0.4em\relax Berlin, BE, DE: Springer,
  1998.

\end{thebibliography}

\begin{IEEEbiography}[{\includegraphics[width=1in,height=1.25in,clip,keepaspectratio]{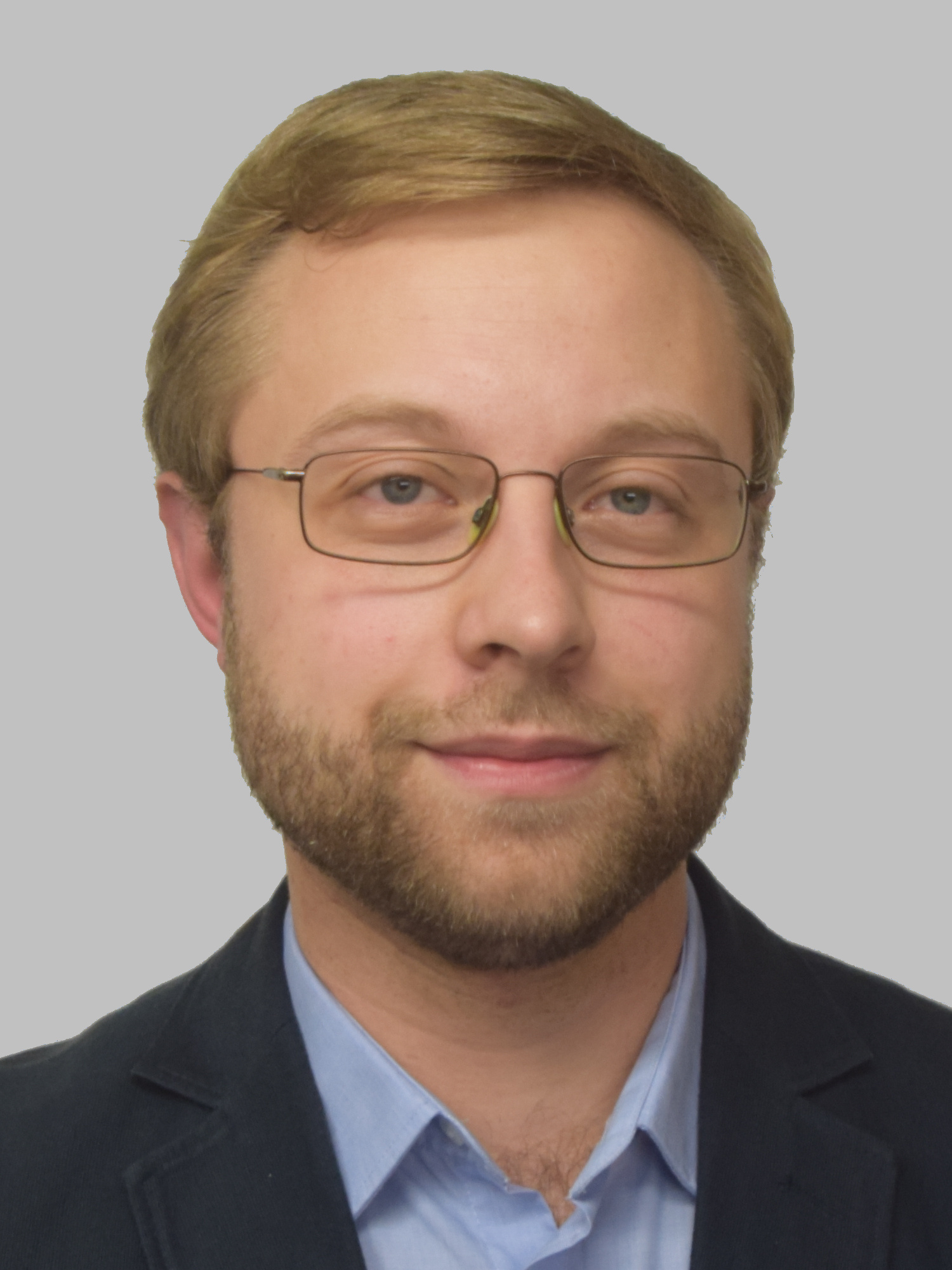}}]{Andreas Martin Kettner}
	(S’15-M’19) received the M.Sc. degree in electrical engineering and information technology from the Swiss Federal Institute of Technology of Zürich (ETHZ), Zürich, Switzerland in 2014, and the Ph.D. degree in power systems engineering from the Swiss Federal Institute of Technology of Lausanne (EPFL), Lausanne, Switzerland in 2019.
	He was a postdoctoral researcher at the Distributed Electrical Systems Laboratory (DESL) of EPFL from 2019 to 2020.
	Since then, he has been working as a project engineer for PSI NEPLAN AG in K{\"u}snacht, Switzerland and continues to collaborate with DESL.
\end{IEEEbiography}

\begin{IEEEbiography}[{\includegraphics[width=1in,height=1.25in,clip,keepaspectratio]{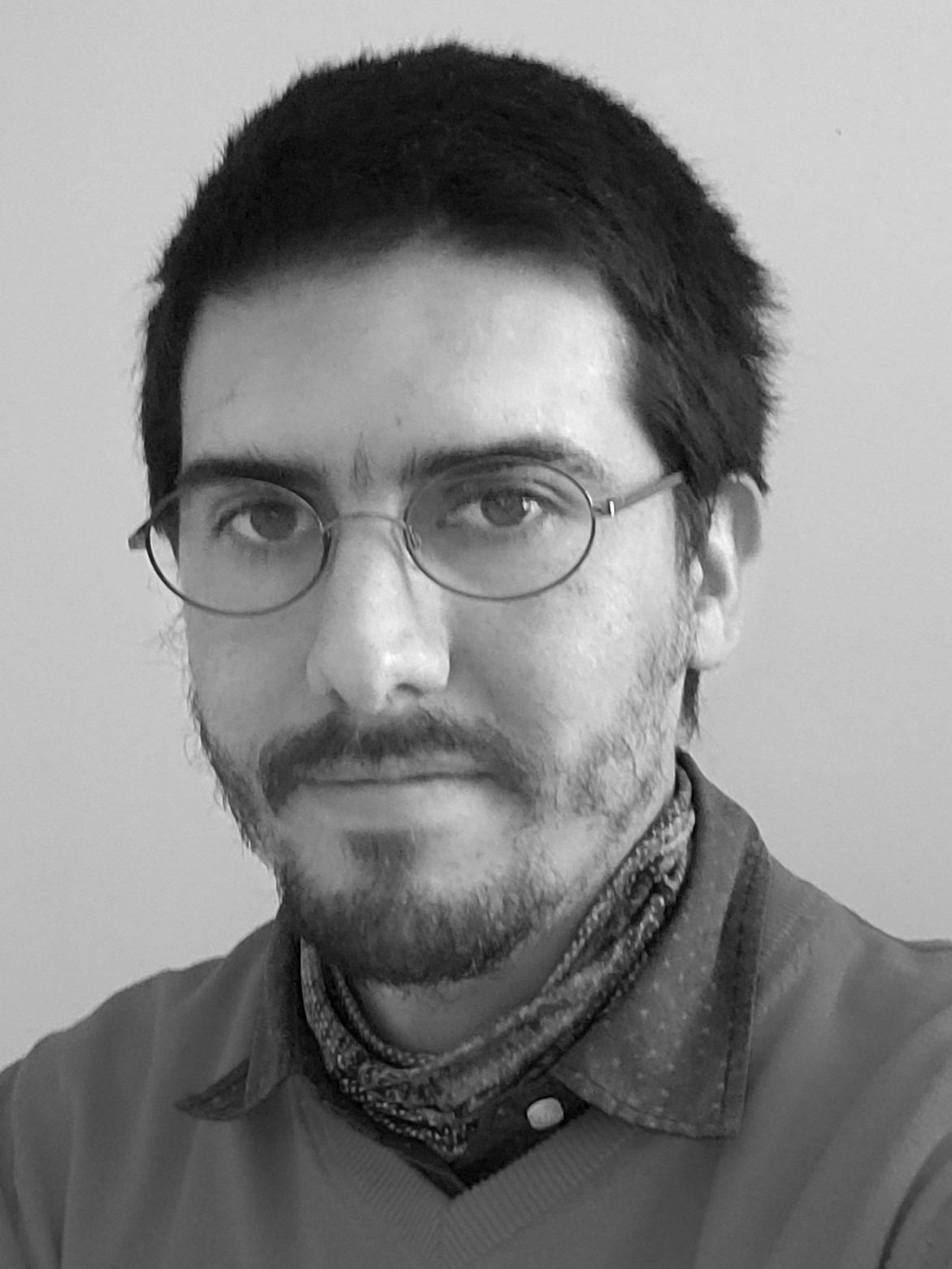}}]{Lorenzo Reyes-Chamorro}
	(S'13-M'16-SM'20) was born in Santiago, Chile in 1984.
    He received the B.Sc. degree in electrical engineering from the University of Chile, Santiago, Chile in 2009 and the Ph.D. degree at the Swiss Federal Institute of Technology of Lausanne (EPFL), Lausanne, Switzerland in 2016.
    He was a postdoctoral fellow at the Distributed Electrical System Laboratory, EPFL from 2016 to 2018. He is currently Assistant Professor in the Institute of Electricity and Electronics and Director of the Innovative Energy Technologies Center (INVENT UACh) at Universidad Austral de Chile, Valdivia, Chile.
\end{IEEEbiography}

\begin{IEEEbiography}[{\includegraphics[width=1in,height=1.25in,clip,keepaspectratio]{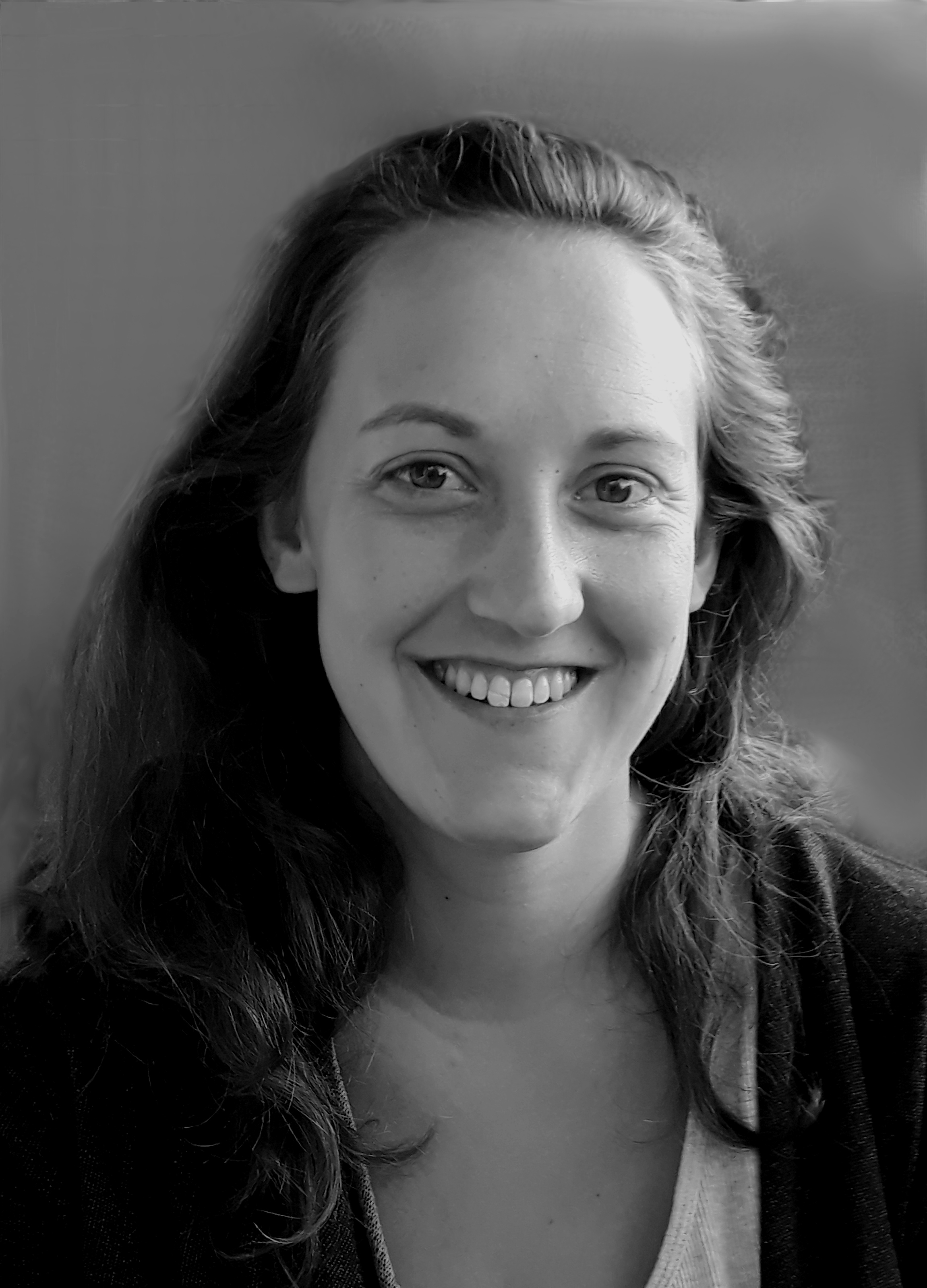}}]{Johanna Kristin Maria Becker}
	(S’19) received the B.Sc. degree in Microsystems Engineering from Freiburg University, Germany in 2015 and the M.Sc. degree in Electrical Engineering from the Swiss Federal Institute of Technology of Lausanne (EPFL), Lausanne, Switzerland in 2019.
    She is currently pursuing a Ph.D. degree at the Distributed Electrical System Laboratory, EPFL, with a focus on robust control and stability assessment of active distribution systems in presence of harmonics.
\end{IEEEbiography}

\begin{IEEEbiography}[{\includegraphics[width=1in,height=1.25in,clip,keepaspectratio]{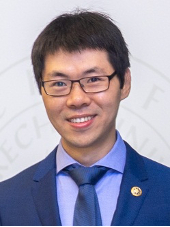}}]{Zhixiang Zou}
	(S’12-M’18-SM’20) received the B.Eng. and Ph.D. degrees in electrical and engineering from Southeast University, Nanjing, China, in 2007 and 2014, respectively, the Dr.-Ing. degree (summa cum laude) from Kiel University, Germany, in 2019. 
	He was an engineer in the State Grid Electric Power Research Institute, Nanjing, China, from 2007 to 2009. 
	He was a research fellow at the Chair of Power Electronics, Kiel University, Germany, from 2014 to 2019. 
	He is now an associate professor in the School of Electrical Engineering at the Southeast University. 
	His research interests include smart transformers, microgrid stability, modeling and control of power converters.
    Dr. Zou serves as an Associate Editor of the IEEE Open Journal of Power Electronics, an Associate Editor of the IEEE Access, an Editor of the International Transactions on Electrical Energy Systems, and an Editor of the Mathematical Problem in Engineering, and a Standing Director of IEEE PES Power System Relaying \& Control Satellite Committee.
\end{IEEEbiography}

\begin{IEEEbiography}[{\includegraphics[width=1in,height=1.25in,clip,keepaspectratio]{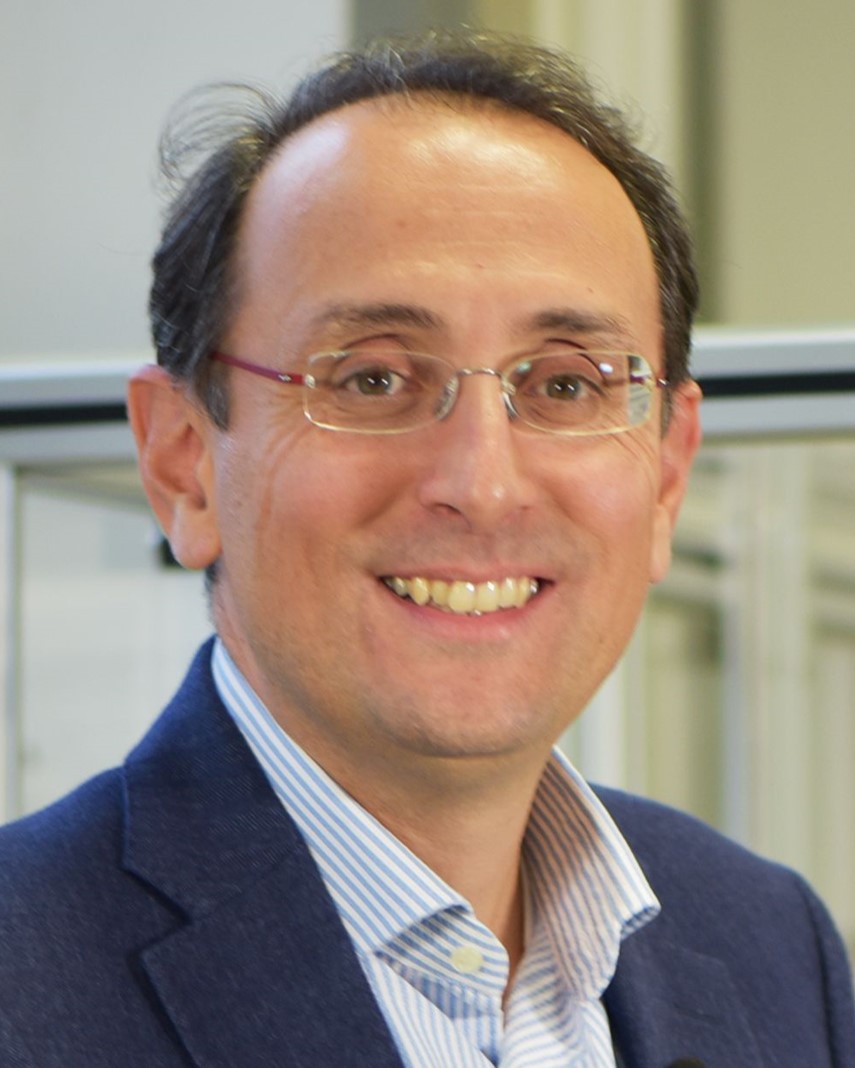}}]{Marco Liserre}
	(S'00-M'02-SM'07-F'13) received the MSc and PhD degree in Electrical Engineering from the Bari Polytechnic, respectively in 1998 and 2002. 
	He has been Associate Professor at Bari Polytechnic and from 2012 Professor in reliable power electronics at Aalborg University (Denmark). 
	From 2013 he is Full Professor and he holds the Chair of Power Electronics at Kiel University (Germany). 
	He has published 500 technical papers (1/3 of them in international peer-reviewed journals) and a book. These works have received more than 35000 citations. 
	Marco Liserre is listed in ISI Thomson report “The world’s most influential scientific minds” from 2014. 
    He has been awarded with an ERC Consolidator Grant for the project “The Highly Efficient And Reliable smart Transformer (HEART), a new Heart for the Electric Distribution System”.
    He is member of IAS, PELS, PES and IES. 
    He has been serving all these societies in different capacities. 
    He has received the IES 2009 Early Career Award, the IES 2011 Anthony J. Hornfeck Service Award, the 2014 Dr. Bimal Bose Energy Systems Award, the 2011 Industrial Electronics Magazine best paper award in 2011 and 2020 and the Third Prize paper award by the Industrial Power Converter Committee at ECCE 2012, 2012, 2017 IEEE PELS Sustainable Energy Systems Technical Achievement Award and the 2018 IEEE-IES Mittelmann Achievement Award.
\end{IEEEbiography}

\begin{IEEEbiography}[{\includegraphics[width=1in,height=1.25in,clip,keepaspectratio]{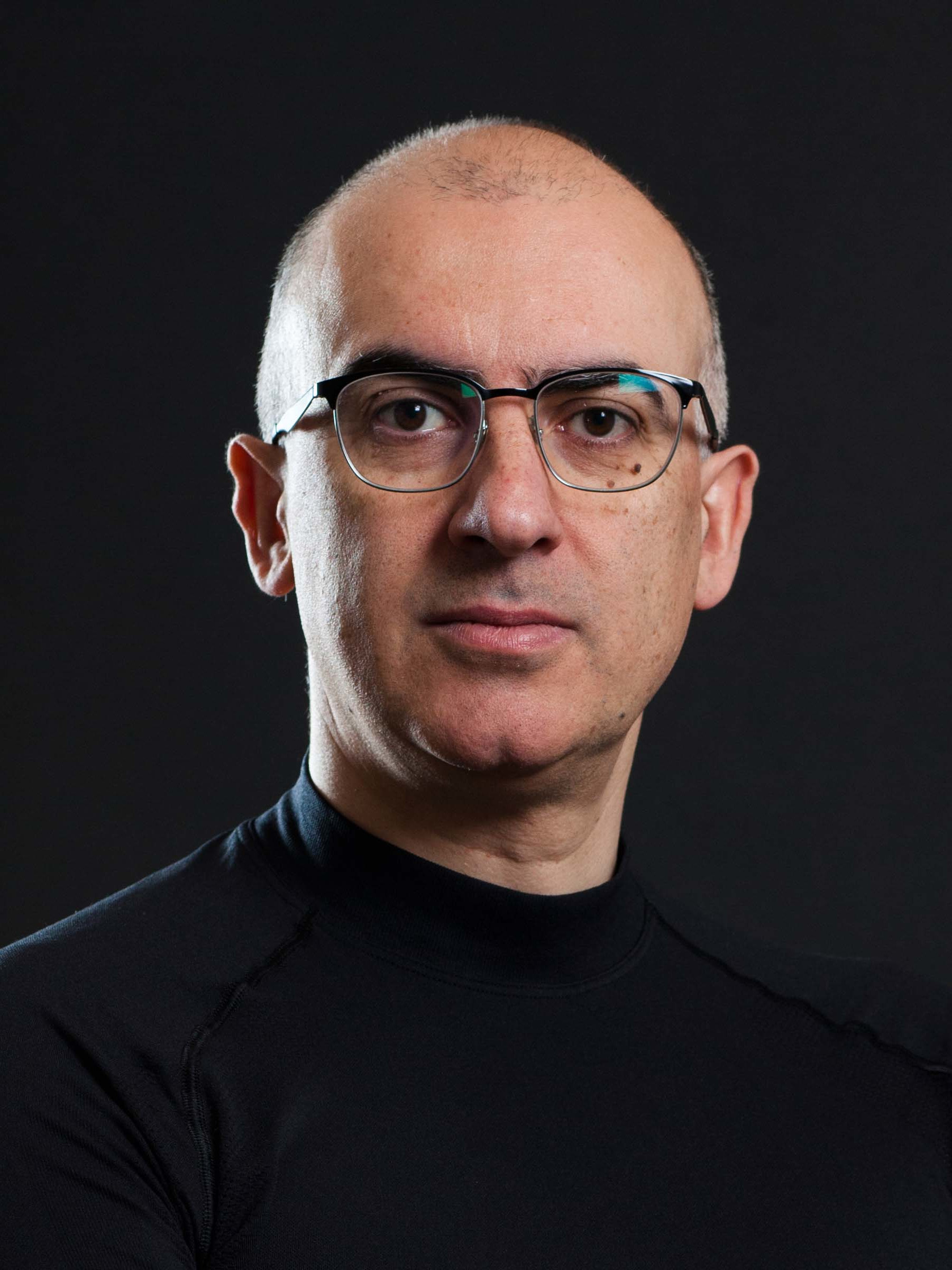}}]{Mario Paolone}
	(M’07–SM’10) received the M.Sc. (Hons.) and Ph.D. degrees in electrical engineering from the University of Bologna, Italy, in 1998 and 2002. 
	In 2005, he was an Assistant Professor in power systems with the University of Bologna, where he was with the Power Systems Laboratory until 2011. 
	Since 2011, he has been with the Swiss Federal Institute of Technology, Lausanne, Switzerland, where he is Full Professor and the Chair of the Distributed Electrical Systems Laboratory. 
	His research interests focus on power systems with reference to real-time monitoring and operational aspects, power system protections, dynamics and transients. 
	Dr. Paolone has authored or co-authored over 300 papers published in mainstream journals and international conferences in the area of energy and power systems that received numerous awards including the IEEE EMC Technical Achievement Award, two IEEE Transactions on EMC best paper awards, the IEEE Power System Dynamic Performance Committee’s prize paper award and the Basil Papadias best paper award at the 2013 IEEE PowerTech. Dr. Paolone was the founder Editor-in-Chief of the Elsevier journal Sustainable Energy, Grids and Networks.
\end{IEEEbiography}

\end{document}